\documentclass[acmjacm]{acmtrans2m}

\newtheorem{theorem}{Theorem}[section]
\newtheorem{lemma}[theorem]{Lemma}
\newdef{definition}[theorem]{Definition}

\newcommand{\process}[3]{\put(0,#2){\Large\ensuremath{p_{#1}}}
                         \put(20,#3){\line(1,0){300}}}

\title{A Unified Theory of Shared Memory Consistency}
\author{ROBERT C. STEINKE and GARY J. NUTT \\
	University of Colorado at Boulder}
\begin{abstract}
The traditional assumption about memory is that a read returns the
value written by the most recent write.  However, in a shared memory
multiprocessor several processes independently and simultaneously
submit reads and writes resulting in a partial order of memory
operations.  In this partial order, the definition of most recent
write may be ambiguous.  Memory consistency models have been developed
to specify what values may be returned by a read given that memory
operations may only be partially ordered.  Before this work,
consistency models were defined independently.  Each model followed a
set of rules which was separate from the rules of every other model.
In our work we have defined a set of four consistency properties.  Any
subset of the four properties yields a set of rules which constitute a
consistency model.  Every consistency model previously described in
the literature can be defined based on our four properties.
Therefore, we present these properties as a unfied theory of shared
memory consistency.

Our unified theory provides several benefits.  First, we claim that
these four properties capture the underlying structure of memory
consistency.  That is, the goal of memory consistency is to ensure
certain declarative properties which can be intuitively understood by
a programmer, and hence allow him or her to write a correct program.
Our unified theory provides a uniform, formal definition of all
previously described consistency models, and in addition some
combinations of properties produce new models that have not yet been
described.  We believe these new models will prove to be useful
because they are based on declarative properties which programmers
desire to be enforced.  Finally, we introduce the idea of selecting a
consistency model as an on-line activity.  Before our work, a shared
memory program would run start to finish under a single consistency
model.  Our unified theory allows the consistency model to change as
the program runs while maintaining a consistent definition of what
values may be returned by each read.
\end{abstract}
\category{}{}{}
\terms{Theory}
\keywords{}
\begin{document}
\begin{bottomstuff}
\end{bottomstuff}
\maketitle

\section{Introduction}

\label{intro}

Shared memory is a powerful abstraction for interprocess
communication.  The concept of shared memory originated from
multiprogramming on uniprocessors and bus-based multiprocessors.  In
these environments there is a simple model of the memory system
enforced in hardware.  The model can be stated as:

\begin{itemize}

\item There is a physical memory cell that represents each variable.
The state of this memory cell is the state of the variable.

\item Memory operations take place sequentially.  They are atomic and
there is a total order on all memory operations.  Read operations
return the current state of the physical memory cell.  Write
operations change the current state of the physical memory cell and
the change becomes observable to all processes simultaneously.

\item The operations of each process take place in the order specified
by its program.

\end{itemize}

These conditions are enforced by the hardware architecture.  In a
multiprogrammed uniprocessor there really is only one process
submitting memory operations at a time.  In a bus-based multiprocessor
with no cache, the bus serves as a serialization mechanism that allows
operations to reach memory sequentially.

For many years these assumptions were implicit and any computer
scientist would tell you, ``That's just how memory works.''  Then two
things happened.  The first is that memory systems in multiprocessors
got more and more
complicated~\cite{dubois90:dep,dubois86:weak,gharachorloo90:release,lenoski90:dash}.
The second is the invention of distributed shared memory (DSM) for the
message-passing
multicomputer~\cite{amza96:treadmarks,bennett90:munin,bennett95:comm,bershad93:midway,bershad91:entry,li86:ivy,li89:ivy}.
Caching and out-of-order instruction dispatching can pose a problem
for multiprocessors.  The hardware of each processor enforces the
restriction that the processor sees its own memory operations in the
order specified by its program, but this does not automatically
protect processors from seeing each other's operations out of order.
DSM provides the illusion of a shared address space on top of hardware
that only supports message passing.  In DSM systems, asynchronous
messages and replicated copies of data can cause the same problems.

These problems led to the concept of consistency models.  A
consistency model is a specification of the allowable behavior of
memory.  It can be seen as a contract between the memory
implementation and the program utilizing
memory~\cite{tanenbaum:dist_os}.  The input to memory is a set of
memory operations (reads and writes) partially ordered by program
order.  The output of memory is the collection of values returned by
all read operations.  A consistency model is a function that maps each
input to a set of allowable outputs.  The memory implementation
guarantees that for any input it will produce some output from the set
of allowable outputs specified by the consistency model.  The program
must be written to work correctly for any output allowed by the
consistency model.  This idea was originally described by Lamport when
he defined sequential consistency~\cite{lamport79:sequential}.  A
sequentially consistent multiprocessor allows conventional reasoning
about the correctness of programs.  Essentially, it allows the
programmer to treat the machine as a multiprogrammed uniprocessor.
Enforcing sequential consistency can be very costly.  Soon weaker
consistency models were discovered that were less expensive in terms
of communication.  Multiprocessors were generally used for large
numerical programs that were already programmed with a constrained
programming style to avoid data race conditions.  With slight
modifications to the programming style, algorithms could still be
written to execute correctly for non-sequentially consistent memory
systems.

With consistency models, the concept of shared memory is no longer
tied to the physical implementation of memory cells.  A programmer can
write a correct program using the abstractions of concurrent processes
and shared memory with little knowledge about the underlying
implementation that will eventually execute the program.  All that the
programmer needs to know is the consistency model enforced by memory.
To give the memory implementor more flexibility for optimization, the
memory might enforce fewer guarantees.  Or to make the programmer's
job easier the memory might enforce more guarantees.  Many choices
have been made along this ease of use to efficient implementation
continuum.  The results are the consistency models described in the
literature~\cite{ahamad91:causal,bershad91:entry,dubois86:weak,gao00:location,gharachorloo90:release,goodman89:processor,herlihy90:linear,hutto90:slow,iftode96:scope,keleher92:lazyrelease,lamport79:sequential,lipton88:pram}

\begin{figure}[tp]

\begin{center}

\begin{picture}(240,85)(0,0)

\put(0,65){\framebox(240,20){Application Program}}
\put(10,50){Consistency Model}
\put(5,40){\vector(0,1){25}}
\put(5,65){\vector(0,-1){25}}
\put(100,55){Read and Write}
\put(100,45){Operations}
\put(95,65){\vector(0,-1){25}}
\put(175,40){\vector(0,1){25}}
\put(180,55){Values Returned}
\put(180,45){by Reads}
\put(0,20){\framebox(240,20){Shared Memory API}}
\put(0,0){\framebox(240, 20){Memory Implementation (Black Box)}}

\end{picture}

\end{center}

\caption{Shared Memory as an API}

\label{api}

\end{figure}
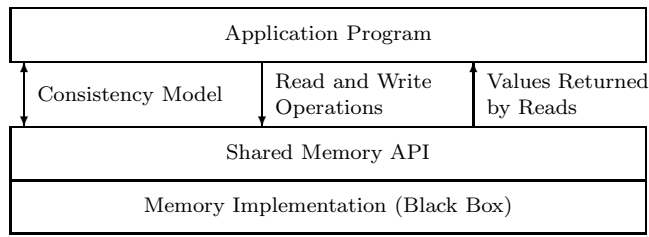

This leads to the idea of shared memory as an application programming
interface (API) as shown in Figure~\ref{api}.  The program and memory
agree on a consistency model.  Then the program executes using the
shared memory API, and the program's processes share information in a
common address space.  No knowledge is needed of the memory
implementation.

This work also introduces the idea of on-line consistency model
transitions.  Prior to this research, the selection of a consistency
model was seen as an off-line activity.  A program would be written to
operate under a particular consistency model, and it would be up to
the user to run the program on a system which supported that
consistency model.  Instead, with consistency model transitions a
program is allowed to select and change the consistency model at
run-time.  The consistency model becomes a tunable parameter to the
shared memory API.  This allows a program to select different
consistency models for different phases of a computation.  This
requires that consistency models be extended with a transition theory
to specify the allowed behavior of the memory system when processing
pending operations submitted under more than one consistency model.

One hypothesis of our work was that every consistency model is
composed of various consistency properties, system-wide conditions
that must be enforced, and that these properties can be combined in
arbitrary ways to produce a lattice of consistency models.  By
defining every consistency model as a set of primitive properties,
transitions between models can be described as the addition or removal
of various properties.  For evaluation and validation, the new
properties proposed in this paper are compared against existing
definitions of consistency models.  Existing consistency models fall
into two classes, Either Non-synchronized or Synchronized models.
Non-synchronized models have uniform consistency restrictions for all
operations.  Synchronized models have special operations (called
synchronization operations) which have greater consistency
restrictions than other operations.  Non-synchronized consistency
models from the literature are simulated by combinations of properties
in the lattice.  Synchronized models have two distinct types of
operations that have different consistency requirements.  Therefore,
synchronized consistency models are simulated by consistency
transitions

The first contribution of this work is the discovery of four
fundamental consistency properties: process order, data order,
write-read-write order, and anti order.  These properties provide
alternate definitions of well known non-synchronized consistency
models and reveal a fundamental structure behind the models.  Every
non-synchronized model described in the literature can be formally
described by some combination of these properties.  The second
contribution of this work is the concept of a consistency lattice.  In
the lattice, each pair of models has a unique least upper bound and a
unique greatest lower bound.  These define the minimum model required
to enforce all conditions of both models, and the maximum set of
conditions enforced by both models respectively.  This lattice allows
simple, direct comparison of models, and is a valuable resource for
any application environment that uses more than one consistency model.
The third contribution of this work is the new consistency models
revealed by the structure of the lattice.  Generating every possible
combination of properties produces five combinations that are well
defined consistency models that have not previously been discovered.
The fourth contribution of this work is a transition theory that can
be used to simulate well known synchronized consistency models.

FIXME Insert roadmap here.

\section{Related Work}

\label{related}

\subsection{Shared Memory}

A common trend in the literature is the development of uniform
frameworks and notation to represent several previously defined
consistency
models~\cite{adve93:data-race-free,adve96:tutorial,bataller97:synch,mosberger93}.
Our unified theory is an improvement over these methods because we
expose the underlying structure of declarative properties enforced by
various models, and we predict new models that have not yet been
discovered.  There are currently two common methods of characterizing
consistency models.  One method is to describe restrictions on the way
in which processes are allowed to issue memory operations which we
will call the ``issue'' method (e.g. see~\cite{adve96:tutorial}.)
Another method is to describe restrictions on the apparent order of
events visible to processes which we will call the ``view'' method
(e.g. see~\cite{bataller97:synch}.)  Adve and
Gharachorloo~\cite{adve96:tutorial} use the ``issue'' method of
defining consistency models.  They identify two conditions that
together will enforce sequential consistency.  They call these the
process order property, and the write atomicity property.

\begin{description}

\item[Process order property] Program order must be maintained among
operations from individual processes.

\item[Write atomicity property] In cache based systems with multiple
copies of a memory location, writes must be atomic.

\end{description}

The first condition can be enforced by having a process not issue an
operation until all previous operations are complete.  Complete means
that a read has returned its value, or a write has been applied and
acknowledged.  The second condition can be enforced by a cache
coherence protocol which does not acknowledge writes until every copy
is updated or invalidated.  Adve and Gharachorloo use this
implementation of sequential consistency as a basis for their
definitions of other consistency models.  Every other model is allowed
to violate some of the restrictions required for sequential
consistency.  Violating a restriction allows for optimization in the
implementation.  They identify five optimizations that may be allowed.

\begin{itemize}

\item Allow a read to be issued before a previous write is complete.

\item Allow a write to be issued before a previous write is complete.

\item Allow a read or write to be issued before a previous read is
complete.

\item Allow a read to view another process' write before the write is
applied everywhere.

\item Allow a read to view one's own write before the write is applied
everywhere.

\end{itemize}

The first optimization combined with the last two result in processor
consistency as it was defined for the DASH
multiprocessor~\cite{lenoski90:dash}.  All five optimizations combined
result in slow consistency~\cite{hutto90:slow} which is used for
non-synchronizing operations in synchronized consistency models such
as release and weak consistency.  For each consistency model, Adve and
Gharachorloo describe a ``safety net'' which would enforce sequential
consistency on top of that model.  These safety nets consist of
replacing certain operations with special purpose synchronization
operations such as test and set or acquire/release.  They also
describe the concept of a programmer centric framework where for any
consistency model a programmer can determine what synchronizations
must be performed for a program to simulate sequential consistency on
top of that model.

The goal of consistency models in this view is to simulate sequential
consistency with an efficient implementation.  The tradeoff is speed
versus complexity exposed to the programmer.  Their work does not
characterize the order of events as seen by any particular process in
a non-sequential execution.  Instead, they characterize what changes a
programmer must make to a program to simulate sequential consistency.
Other work taking this view has been done to present an efficient,
sequentially consistent interface to the programmer through
instruction level parallelism and speculative
execution~\cite{gniady99:ilp,ranganathan97:ilp,ranganathan97:spec}.
The logic being that speculative rollback will generally occur in
situations where the processes would be waiting on synchronization
operations anyway so little time would actually be lost.

We believe that using weaker consistency models soley to simulate
sequential consistency with an efficient implementation should not be
the only goal of shared memory research.  Our work is based on the
idea of declarative consistency properties weaker than sequential
consistency, but still intuitively useful to programmers.  Therefore,
we found the formalisms of the ``issue'' method less useful to us.

An alternative to the ``issue'' method is the ``view'' method where
each process has a view of the order of events in the system.  For
example, PRAM consistency~\cite{lipton88:pram} states that each
process must see all operations to occur in an order that respects
program order, but different processes may see different orders.  This
essentially places restrictions on when operations may become visible
to other processes, and not on when they may be issued.  For our
purposes, the view method of defining consistency models is most
appropriate.  What matters is the possible orders of events from the
process' (programmer's) point of view.  The programmer does not care
how the shared memory is implemented.  If two different
implementations produce the same set of possible views they should be
considered equivalent.  For this reason, our work uses view based
definitions of consistency models.  We believe they are more
independent of implementation details.  Several surveys of view based
definitions have been presented in the
literature~\cite{bataller97:synch,mosberger93,tanenbaum:dist_os}.
These view based definitions are presented in
Subsections~\ref{non-synch} and~\ref{synch}.

The only prior comparison in the literature of the issue and view
methods is by Mustaque Ahamad, et. al.~\cite{ahamad92:processor}.  In
their paper they compare Goodman's definition of processor consistency
(which is view based) to the DASH definition (which is issue based.)
Their conclusion was that both definitions are weaker than sequential
consistency, and stronger than both PRAM and cache consistency.  This
is the strength relationship commonly understood for processor
consistency, and the two models have often been considered equivalent.
However, Ahamad, et. al. showed that the two definitions are not
equivalent, and are in fact incomparable.  This showed that it is not
trivial to compare consistency models defined under the two
formalisms.  More work relating the two formalisms is needed.
However, this paper concentrates on view based definitions.
Generally, issue based definitions have a view based definition that
is analogous.

The most closely related work to this paper is the Mume
project~\cite{bataller98:mume} which specifies three consistency
properties (orderings): total order, total order with mutual
exclusion, and causal order.  The Mume project showed that these
orderings can be used to provide an alternative and equivalent
specification of existing consistency models.  However, unlike our
work, there is no notion of combining properties in arbitrary ways to
produce a lattice of consistency models, or of consistency transitions
within that lattice.

\subsection{Consistency Model Definitions}

\label{non-synch}

Leslie Lamport defined sequential
consistency~\cite{lamport79:sequential}:

\begin{definition}

A multiprocessor is \emph{Sequentially Consistent} if the result of
any execution is the same as if the operations of all the processors
were executed in some sequential order, and the operations of each
individual processor appear in this sequence in the order specified by
its program.

\end{definition}

Lamport also gave two implementation requirements which, if met, would
enforce sequential consistency.

\begin{description}

\item[R1] Each processor issues memory requests in the order specified
by its program.

\item[R2] Memory requests from all processors issued to an individual
memory module are serviced from a single FIFO queue.  Issuing a memory
request consists of entering the request on this queue.

\end{description}

Linearizability~\cite{herlihy90:linear} also called atomic
memory~\cite{lamport86:atomic} is essentially sequential consistency
with a real-time constraint.  Each operation is given a begin time and
end time in reference to a global Newtonian clock.  For an execution
to be linearizable, it must be sequentially consistent, and the
sequential total order must correspond to an order realizable by
placing each operation at a single point in time between its begin and
end times.  Essentially, if two operations' time spans do not overlap
they cannot be re-ordered even in the absence of any other dependency.
Even though linearizability is stronger, sequential consistency is the
strongest consistency model used in
practice~\cite{adve96:tutorial,tanenbaum:dist_os}.  Sequential
consistency is considered strong enough for conventional reasoning
about the correctness of shared memory programs.

Lipton and Sandberg defined PRAM (Pipelined RAM)
consistency~\cite{lipton88:pram}, and Goodman defined cache
consistency~\cite{goodman89:processor}:

\begin{definition}

A multiprocessor is \emph{PRAM Consistent} if writes performed by a
single process are seen by all other processes in the order in which
they were issued, but writes from different processes may be seen in
different orders by different processes.

\end{definition}

\begin{definition}

A multiprocessor is \emph{Cache Consistent} if all writes to the same
memory location are performed in some sequential order.

\end{definition}

In the same paper Goodman defined processor consistency.

\begin{definition}

A multiprocessor is \emph{Processor Consistent} if it is PRAM
consistent and writes to the same memory location are seen in the same
sequential order by all processes.

\end{definition}

\begin{figure}[tp]

\begin{center}

\begin{picture}(320,40)(0,0)

\process{1}{25}{20} \put(60,23){$(w,p_1,x,1)$}
\put(160,23){$(r,p_1,y,\bot)$} \process{2}{5}{0}
\put(100,3){$(w,p_2,y,2)$} \put(200,3){$(r,p_2,x,\bot)$}

\end{picture}

\end{center}

\caption{An execution that is processor, but not sequentially
consistent.}

\label{processor}

\end{figure}
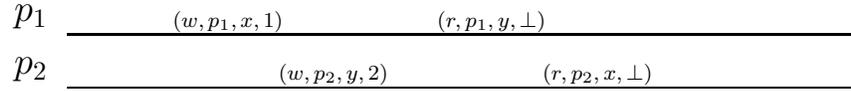

One consistency model is said to be stronger than another if every
condition required by the weaker model is also required by the
stronger one.  Thus, a stronger consistency model has a more highly
constrained behavior than a weaker one.  By considering the
definitions, note that sequential consistency is strictly stronger
than processor consistency which is strictly stronger than both PRAM
and cache consistency.  However, PRAM and cache consistency are
incomparable.  PRAM and cache consistency are very similar to
Lamport's conditions R1 and R2, enforcing R1 and R2 enforces
sequential consistency, processor consistency enforces PRAM
consistency and cache consistency, but processor consistency is weaker
than sequential consistency.  How can this be?

Consider Figure~\ref{processor}.  In this figure time proceeds from
left to right, and variables are assumed to have an initial value of
$\bot$.  Process $p_1$ writes to $x$, and then reads from $y$.
Likewise, process $p_2$ writes to $y$, and then reads from $x$.  Both
processes read the initial value of the variable instead of each
other's write.  Both processes perceive that their write went first so
the execution is not sequential.  However, it is processor consistent.
There is only one write by $p_1$ and one by $p_2$ so it is trivially
PRAM consistent.  There is only one write to $x$ and one write to $y$
so it is trivially cache consistent.  This example demonstrates how
processor consistency is weaker than sequential consistency.  Writes
by different processes to different variables may be seen to occur in
different orders.

The question remains, does the execution in Figure~\ref{processor}
satisfy R1 and R2?  The answer is no because R2 requires that read
operations be placed in the queue along with write operations.
Neither process can place its read operation in the queue until its
write operation has been placed in the queue so at least one of the
processes must read the other's write.  On the other hand, processor
consistency only requires that write operations become visible in the
correct order.  The write operations can be pending while each process
does its read, and then the write operations are applied in the
correct order.

Causal memory~\cite{ahamad91:causal} is a consistency model drawn from
Lamport's concept of potential causality~\cite{lamport78:causality}.
Causal memory is weaker than sequential consistency, stronger than
PRAM, and incomparable to processor and cache consistency.  It was
defined by Ahamad, et. al. as:

\begin{definition}

A multiprocessor is \emph{Causally Consistent} if for each process the
operations of that process plus all writes known to that process
appear to that process to occur in a total order that respects
potential causality.  Potential causality is as defined by
Lamport~\cite{lamport78:causality} with writes interpreted as sends
and reads interpreted as receives.

\end{definition}

Slow consistency~\cite{hutto90:slow} is weaker than both PRAM and
cache consistency.

\begin{definition}

A multiprocessor is \emph{slow consistent} if reads must return some
value that has been previously written to the location being read.
Once a value has been read, no earlier writes to that location (by the
processor that wrote the value read) can be returned.  Writes by a
process must be immediately visible to itself.

\end{definition}

Local consistency~\cite{bataller97:synch} refers to the weakest
consistency model for shared memory.

\begin{definition}

A multiprocessor is \emph{Locally Consistent} if each process' own
operations appear to occur in the order specified by its program.
There is no restriction on the order in which writes by other
processes appear to occur, and different processes may see different
orders.

\end{definition}

It is important to note that every consistency model is stronger than
local consistency and weaker than sequential consistency which is
weaker than linearizability~\cite{herlihy90:linear}.  This fact
implies that consistency models could be placed in a lattice.

\subsection{Synchronized Consistency Models}

\label{synch}

Some consistency models include explicit synchronization actions which
are treated differently than ordinary memory operations.
Synchronization operations are processed at a high level of
consistency, usually sequential consistency.  Ordinary operations are
processed at a low level of consistency, usually slow consistency, but
the presence of synchronization operations places additional ordering
restrictions on ordinary operations.  Dubois, et. al. defined weak
consistency~\cite{dubois86:weak}.

\begin{definition}

A multiprocessor is \emph{Weak Consistent} if:

\begin{enumerate}

\item Accesses to global synchronizing variables are strongly ordered
[sequentially consistent].

\item No access to a synchronizing variable is issued in a processor
before all previous global data accesses have been performed.

\item No access to global data is issued by a processor before a
previous access to a synchronizing variable has been performed.

\end{enumerate}

\end{definition}

An ordinary operation is issued either before or after a
synchronization operation.  All processes must see the ordinary
operation occur in this order with respect to the synchronization
operation.  This provides a sufficient programming environment for
constructs such as critical sections and barriers.  For example, a
barrier is defined to be a synchronization operation, and all
operations issued before the barrier must appear to occur before the
barrier.  However, this condition is sometimes stronger than
necessary.  Synchronizing operations can be used just to import
information, as with the acquiring of a lock, or just to export
information, as with the release of a lock.  Taking advantage of this
as an opportunity for optimization leads to a different consistency
model called release consistency~\cite{gharachorloo90:release}.

\begin{definition}

A multiprocessor is \emph{Release Consistent} if:

\begin{enumerate}

\item Before an ordinary LOAD or STORE access is allowed to perform
with respect to any other processor, all previous \emph{acquire}
accesses must be performed.

\item Before a \emph{release} access is allowed to perform with
respect to any other processor, all previous ordinary LOAD and STORE
accesses must be performed.

\item \emph{Special accesses} [including acquire and release] are
sequentially consistent with respect to one another.

\end{enumerate}

\end{definition}

A process performs an acquire to get up to date information.  Only
that process is guaranteed to be up to date, and then only up to the
point of the latest release on every other process.  A different
implementation called lazy release
consistency~\cite{keleher92:lazyrelease} enforces the same consistency
model, but sends updates as late as possible.  The distinction between
release and weak consistency is that release forces the program to
give more detailed instructions on what must be up to date at a
synchronization.  This trend is continued with entry
consistency~\cite{bershad91:entry} and scope
consistency~\cite{iftode96:scope}.  In entry
consistency~\cite{bershad91:entry} each synchronization variable is
associated with one or more ordinary variables.  Acquires and releases
only bring up to date those ordinary variables associated with a
particular synchronization variable.  In scope
consistency~\cite{iftode96:scope} this set of variables is not static,
but rather any ordinary variables accessed between an acquire and
release of a synchronization variable must be brought up to date to
the point of the release on all subsequent acquires of the same
synchronization variable.

A final synchronized model called location
consistency~\cite{gao00:location} is significantly different.
location consistency is similar to entry consistency in that each
ordinary variable is associated with a synchronization variable, and a
release or acquire is ordered with an ordinary operation if their
variables are associated.  However, location consistency is different
in that it allows the state of a variable to be a partial order, and
not a total order.

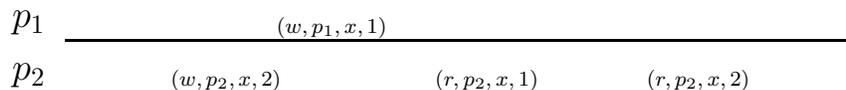
\begin{figure}[tp]

\begin{center}

\begin{picture}(320,40)(0,0)

\process{1}{25}{20}
\put(100,23){$(w,p_1,x,1)$}
\process{2}{5}{0}
\put(60,3){$(w,p_2,x,2)$}
\put(160,3){$(r,p_2,x,1)$}
\put(240,3){$(r,p_2,x,2)$}

\end{picture}

\end{center}

\caption{An execution that is location, but not entry consistent.}

\label{location}

\end{figure}

For example, in Figure~\ref{location} two processes both write to the
variable $x$.  In entry consistency, the order of these two writes is
undefined.  They could be seen to occur in either order, and two
different processes do not have to agree on the order.  However, there
is an implicit assumption that for a single process the two operations
occur in some order, and the second one overwrites the first.  So,
when $p_2$ reads $1$ from $x$ one can deduce that the order seen by
$p_2$ is:

\begin{quote}

$(w,p_2,x,2)<(w,p_1,x,1)<(r,p_2,x,1)$

\end{quote}

Therefore, $p_2$ will never again read from $(w,p_2,x,2)$ because it
has been overwritten.  The operation $(r,p_2,x,2)$ violates entry
consistency, but not location consistency.  Location consistency
assumes that each process sees a partial order of writes, and any read
can return the value of any write that is not dominated by another
write.  Writes are only ordered when they are by the same process, or
when they are separated by a release-acquire pair.  Therefore, under
location consistency $p_2$ can continue forever alternately reading
the values $1$ and $2$ from $x$ barring further write, acquire, or
release operations.  The purpose of location consistency is that if a
program separates every pair of competing writes with a
release-acquire pair (called a data-race-free program) then it is
equivalent to entry consistency, but still might be able to take
advantage of the location model for efficiency optimizations.

\section{A Formalism for Shared Memory Consistency Models}

\label{formalism}

This section presents formal, declarative definitions of the well
known consistency models introduced in Section~\ref{related}.  When a
shared memory system satisfies a particular consistency model it must
produce only executions acceptable to that model.  In this way, a
consistency model can be thought of as a criteria to accept or reject
program executions.  Therefore, a model can be defined by specifying
its set of accepted executions.  This is the technique we will use in
the rest of the paper.

In ``view'' based definitions of consistency models, memory operations
must \emph{appear} to be processed in a certain order.  For example,
under sequential consistency, there must appear to be a single total
order on all operations.  Under Cache consistency, there must appear
to be a total order on the operations to each variable.  Each process
sees, through its read operations, a particular order of events in the
memory system.  However, each process has limited information because
it may not read every write.  Therefore, there could be many orders of
events that would be consistent with the values returned by a process'
reads.  If any of these orders satisfies a consistency model then the
process cannot prove that the memory system violated that model.  If
some acceptable order exists for every process then the execution must
be accepted.  The formalism used in this section is defined in the
appendix and is taken from~\cite{ahamad92:processor,bataller97:synch}.

\begin{theorem}

An execution is \emph{Sequentially Consistent} iff

\begin{quote}

$\exists$ SerialView$(<_{PO})$

\end{quote}

\end{theorem}

\begin{quotation}

For proof see~\cite{bataller97:synch}.

\end{quotation}

This restatement of sequential consistency corresponds very closely to
the original definition of sequential consistency.  There exists a
serial view (total order) on all operations that respects $<_{PO}$
(the process order of every process.)  The actual execution may not
have occurred in this order, but the values returned by the reads are
exactly the same as the values that would have been returned had this
been the execution order.  Therefore, no process external to the
memory system can prove that the execution did not actually happen in
this order.  In Figure~\ref{executions}(a) the given total order
qualifies as the serial view proving that the execution is
sequentially consistent.  In Figure~\ref{executions}(b) it is easy to
see that no such view could be constructed.

\begin{theorem}

An execution is \emph{PRAM Consistent} iff

\begin{quote}

$\forall_{i\in P}\exists$ SerialView$(<_{PO}|(*,i,*,*)\cup(w,*,*,*))$

\end{quote}

\end{theorem}

\begin{quotation}

For proof see~\cite{bataller97:synch}.

\end{quotation}

PRAM consistency requires that each process see a view that is
consistent with the process order for all processes, but not all
processes must see the same view.  The operations visible to each
process are its own reads and all writes.  For this reason the view of
process $i$ is restricted to $(*,i,*,*)$, all of its own operations,
and $(w,*,*,*)$, all writes.  If a serial view conforming to process
order can be constructed for this subset of operations then this
process cannot argue that the memory system has violated PRAM.  If
such a view can be constructed for every process then no external
observer can argue that the memory system has violated PRAM.

\begin{theorem}

An execution is \emph{Cache Consistent} iff

\begin{quote}

$\forall_{x\in V}\exists$ SerialView$(<_{PO}|(*,*,x,*))$

\end{quote}

\end{theorem}

\begin{quotation}

For proof see~\cite{bataller97:synch}.

\end{quotation}

Cache consistency requires that for the operations on each variable,
$x$, there is a serial view that respects process order.  The views
that must be constructed to satisfy the above definition are exactly
the total orders required for the original definition.

\begin{figure}[tp]

\begin{center}

\begin{tabular}{l|l}

$(w,p_1,x,1)<_{PO}(r,p_1,x,2)$  & $(w,p_1,x,1)<_{PO}(w,p_1,y,2)$    \\
$(w,p_2,x,2)<_{PO}(r,p_2,x,1)$  & $(r,p_2,y,2)<_{PO}(r,p_2,x,\bot)$ \\
$(w,p_1,x,1)\mapsto(r,p_2,x,1)$ & $(w,p_1,y,2)\mapsto(r,p_2,y,2)$   \\
$(w,p_2,x,2)\mapsto(r,p_1,x,2)$ 
                      & $(w,\epsilon,x,\bot)\mapsto(r,p_2,x,\bot)$  \\
                                                                    \\
\multicolumn{1}{c}{(a)}    & \multicolumn{1}{c}{(b)}

\end{tabular}

\end{center}

\caption{Examples for PRAM and Cache Consistency}

\label{PRAM}

\end{figure}

Consider Figure~\ref{PRAM}.  The sets $P$, $V$, and $O$ and the
initial writes can usually be deduced from the descriptions of process
order and writes-to order.  For this reason they will be omitted in
this and further examples unless required for clarity.  In
Figure~\ref{PRAM}(a), both processes write and then read $x$, and both
read the other's write.  This execution can be shown to be PRAM
consistent by the following views.

\begin{quote}

$p_1:(w,p_1,x,1)<_{p_1}(w,p_2,x,2)<_{p_1}(r,p_1,x,2)$ \\
$p_2:(w,p_2,x,2)<_{p_2}(w,p_1,x,1)<_{p_2}(r,p_2,x,1)$

\end{quote}

This execution is not sequential.  One would have to add $(r,p_2,x,1)$
to $<_{p_1}$, or $(r,p_1,x,2)$ to $<_{p_2}$.  In $<_{p_1}$,
$(r,p_2,x,1)$ cannot come before $(w,p_2,x,2)$ because that would
violate process order.  It also cannot come after $(w,p_2,x,2)$
because then it would be after, but not reading from, $(w,p_2,x,2)$
which would violate the serial property.  A similar argument can be
made for $<_{p_2}$.  No single view can satisfy both processes so the
execution is not sequentially consistent.

In Figure~\ref{PRAM}(b) process 1 writes to both $x$ and $y$ while
process 2 reads both $x$ and $y$.  Process 2 reads process 1's second
write to $y$ and the initial value of $x$.  This execution can be
shown to be Cache consistent by the following views.  Note, the
initial writes must be accounted for in all views, but are omitted in
examples where their placement is trivial.  $(w,\epsilon,x,\bot)$ is
shown in $<_x$ because it's value is later read.

\begin{quote}

$x:(w,\epsilon,x,\bot)<_x(r,p_2,x,\bot)<_x(w,p_1,x,1)$ \\
$y:(w,p_1,y,2)<_y(r,p_2,y,2)$

\end{quote}

Figure~\ref{PRAM}(b) is not sequentially consistent.  In a view with
every operation, $(w,p_1,x,1)$ would have to come before $(w,p_1,y,2)$
by process order.  $(w,p_1,y,2)$ would have to come before
$(r,p_2,y,2)$ for the view to be serial.  $(r,p_2,y,2)$ would have to
come before $(r,p_2,x,\bot)$ by process order.  This implies
$(r,p_2,x,\bot)$ would come after $(w,p_1,x,1)$ but read from the
initial write so the view could not be serial.

Also, ~\ref{PRAM}(a) is not Cache consistent, and ~\ref{PRAM}(b) is
not PRAM consistent.  In ~\ref{PRAM}(a) all operations are on the same
variable so there would need to be a serial view on all operations.
In disproving sequential consistency we have already shown this is
impossible.  For ~\ref{PRAM}(b) to be PRAM consistent the view
$<_{p_2}$ would need to be constructed containing all of $p_1$'s
writes, and all of $p_2$'s operations.  This would include all of the
operations which have likewise been shown to be impossible.

\begin{theorem}

An execution $\alpha$ is \emph{Processor Consistent} iff

\begin{quote}

$\forall_{x\in V}\exists<_x=$SerialView$(<_{PO}|(*,*,x,*))$, and\\
$\forall_{i\in P}\exists$ SerialView$((\cup_{x\in
V}<_x)\bigcup<_{PO}|(*,i,*,*)\cup(w,*,*,*))$

\end{quote}

\label{def-processor}

\end{theorem}

\begin{quotation}

For proof see~\cite{bataller97:synch}.

\end{quotation}

This restatement says that Processor consistency requires PRAM and
cache consistency.  It also requires that the PRAM and cache views be
mutually consistent.  The views that satisfy PRAM must conform not
only to the process order, but to the view order of every variable
enforced by cache consistency.  This is equivalent to Goodman's
definition of processor consistency.

\begin{definition}

The \emph{Causal Relation}, $<_{CR}$,

\begin{quote}

\begin{tabbing}
1234\=\kill

$\forall_{o_i,o_j\in O}\ o_i<_{CR}o_j$ iff\\
\>$o_i<_{PO}o_j,$ or\\
\>$o_i\mapsto o_j,$ or\\
\>$\exists\ o_k\in O$ such that $o_i<_{CR}o_k<_{CR}o_j$

\end{tabbing}

\end{quote}

\end{definition}

\begin{theorem}

An execution $\alpha$ is \emph{Causally Consistent} iff

\begin{quote}

$\forall_{i\in P}\exists$ SerialView$(<_{CR}|(*,i,*,*)\cup(w,*,*,*))$

\end{quote}

\end{theorem}

\begin{quotation}

For proof see~\cite{bataller97:synch}.

\end{quotation}

\begin{theorem}

An execution $\alpha$ is \emph{Slow Consistent} iff

\begin{quote}

$\forall_{i\in P,x\in V}\exists$
SerialView$(<_{PO}|(*,i,x,*)\cup(w,*,x,*))$

\end{quote}

\end{theorem}

\begin{quotation}

For proof see~\cite{bataller97:synch}.

\end{quotation}

\begin{theorem}

An execution $\alpha$ is \emph{Locally Consistent} iff

\begin{quote}

$\forall_{i\in P}\exists$
SerialView$(<_{iLocal}|(*,i,*,*)\cup(w,*,*,*))$

\end{quote}

\end{theorem}

\begin{quotation}

For proof see~\cite{bataller97:synch}.

\end{quotation}

\subsection{Synchronized Consistency Models}

Synchronized consistency models require additional definitions. First
of all, operations are divided into two types, ordinary and
synchronization operations.  In some models such as weak consistency,
reads and writes are merely designated as synchronization operations.
In other models such as release consistency, synchronization
operations are new types of operations, acquire and release.  In
either case, the operation type $s$ is used to designate
synchronization operations.  For example, $(s,*,*,*)$ designates the
set of all synchronization operations whether those are read, write,
acquire, or release.  Also, we need to explicitly state that the
writes-to relation is defined on synchronization operations.  For this
purpose, acquires are treated as reads, and releases are treated as
writes.  Essentially, synchronization operations must be aware of
which acquire corresponds to which release.  Defining the writes-to
relation in this way allows the existing definition of serial view to
be used for this purpose.  Finally, for each synchronized consistency
model, certain ordinary operations must come before or after certain
synchronization operations.

\begin{definition}

$D_-(s)$ denotes the set of ordinary operations that must come before
synchronization operation $s$.  $D_+(s)$ denotes the set of ordinary
operations that must come after synchronization operation $s$.  $<_D$
denotes the relation:

\begin{quote}

$\forall_{o\in D_-(s)}o<_Ds\cup \forall_{o\in D_+(s)}s<_Do$

\end{quote}

\end{definition}

Synchronized consistency models support different consistency for
ordinary operations than synchronization operations.  For some models,
ordinary operations are processed under slow consistency, and for some
models under cache consistency.  The authors
of~\cite{bataller97:synch} argue that this distinction is not a
significant design feature, but rather was primarily an artifact of
the implementation for which each model was originally defined.  They
present formal definitions of all models assuming that ordinary
operations are processed under slow consistency.  Synchronization
operations are generally processed under sequential consistency,
although a variation of release consistency was presented where
synchronization operations were processed under processor consistency.
Below, we assume synchronization operations obey seqential consistency
and ordinary operations obey slow consistency.  Variations will be
dealt with in the section on consistency transitions (see
Section~\ref{transitions}.)

Every synchronized consistency model obeys the following condition.
The only difference between models is in the definition of $D_-(s)$
and $D_+(s)$.

\begin{definition}

\label{synch-def}

For a given definition of $<_D$, an execution is \emph{synchronized
model consistent} iff

\begin{quote}

$\exists<_{seq}=$SerialView$(<_{PO}|(s,*,*,*))$, and\\ $<_S=$the
transitive closure of $<_D\cup<_{seq}$, and\\ $\forall_{i\in P, x\in
V}\exists$ SerialView$(<_S\cup<_{PO}|(*,i,x,*)\cup(w,*,x,*))$

\end{quote}

\end{definition}

Definition~\ref{synch-def} says that a sequential order exists on all
synchronization operations.  The per-process, per-variable views
required by slow consistency exist.  And the slow consistent views
respect the transitive closure of the ordering $<_D$ and the
sequential order of synchronization operations.  We will now discuss
the differences between various consistency models.

In weak consistency~\cite{dubois86:weak} there is only one
synchronizing variable, and there is no distinction between acquire
and release types of synchronizing operations.  $D_+(s)$ orders after
$s$ any operation ordered after it by process order.  $D_-(s)$ orders
before $s$ any operation ordered before it by process order.

For release consistency~\cite{gharachorloo90:release} there is only
one synchronizing variable, but the distinction is made between
acquire and release types of synchronizing operations.  $D_+(acquire)$
orders after $acquire$ any operation ordered after it by process
order.  $D_-(release)$ orders before $release$ any operation ordered
before it by process order.

Lazy release consistency~\cite{keleher92:lazyrelease} does not force
operations before a release to be ordered before that release, but
they must be ordered before any subsequent acquire.  There is only one
synchronizing variable.  $D_+(acquire)$ orders after $acquire$ any
operation ordered after it by process order.  $D_-(acquire)$ orders
before $acquire$ any ordinary operation where there exists $release<_S
acquire$ such that the ordinary operation is ordered before $release$
by process order.  No ordinary operations are directly ordered with
any release.

In entry consistency~\cite{bershad91:entry} there can be more than one
synchronization variable.  Each ordinary variable is associated with a
synchronization variable.  An ordinary operation is ordered with a
synchronization operation in the same way it would by release
consistency if and only if their variables are associated.

In scope consistency~\cite{iftode96:scope} there can be more than one
synchronization variable.  An ordinary operation is ordered with a
synchronization operation in the same way it would be by release
consistency if and only if there is no other synchronization operation
to the same variable ordered between them by process order.
Essentially, ordinary operations are only ordered with respect to the
most recent acquire and the next release to each synchronization
variable.

Location consistency~\cite{gao00:location} is different, but we will
present it in a formalism as close as possible to that used for the
other models.  One important difference is that in location
consistency synchronization operations are defined to provide a mutual
exclusion function.  If one process performs an acquire, then no other
process may successfully perform an acquire until the first process
performs a release.  All subsequent acquires are ordered after that
release.  This provides control dependencies in addition to the data
dependencies enforced by the consistency model.  This exposes a
fundamental difference of opinion about what the job of a consistency
model should be.  For example, under release consistency there is
nothing to prevent two processes from both performing acquires and
concurrently writing to the same variable.  Release consistency
specifies formally what data dependencies must be preserved by the
memory system in that situation, i.e. the writes are unordered and can
be seen in different orders by different processes.  If the program
truly needs mutual exclusion it can be included in the program code as
a locking algorithm that works correctly under release
consistency~\cite{gharachorloo90:release}.

Most synchronized consistency models were written in two parts, the
consistency model itself, and a programming paradigm such as properly
labeled~\cite{gharachorloo90:release} or
data-race-free~\cite{adve93:data-race-free} programs.  The guarantee
provided is that a program that obeys the programming paradigm
executed on the consistency model will simulate sequential
consistency.  The authors of release consistency expected that it
would be used in conjunction with control flow constructs in the
program to simulate sequential consistency, but they did not directly
embed the control flow into the consistency model.  Instead they
allowed the programmer to choose the appropriate control flow
constructs.  They also acknowledged that some programmers may not want
to simulate sequential consistency, but rather deal directly with the
semantics of release consistency.

Control dependencies should be dealt with in the programming paradigm,
and not the consistency model itself.  It is unnecessary for a
consistency model to force the programmer to use a particular control
flow paradigm like mutual exclusion.  The consistency model should
only describe data dependencies.  For any sequence of submitted
operations the model gives the set of possible outcomes.  It is not
the job of the consistency model to restrict the sequences of
operations that are allowed to be submitted.  Any control dependencies
can be independently enforced in the program.  If the programmer
really wants mutual exclusion the consistency model does not prevent
this.  This does not necessarily even make the programmer's job any
harder as control flow constructs can be implemented in libraries of
locking and barrier primitives.

Synchronization operations in location consistency are similar to
entry consistency in that they are tagged with a variable, and only
enforce dependencies with ordinary operations to that variable.  The
mutual exclusion assumption stated above requires that there is a
total order on all synchronization operations to each variable so
location consistency enforces at least cache consistency on
synchronization operations.  However, the description of location
consistency~\cite{gao00:location} does not specifically say that
synchronization operations must obey sequential consistency.  There is
no example in the paper with synchronization operations to more than
one variable so it is difficult to say whether the authors intended
synchronization operations to be sequentially consistent, or merely
cache consistent.  For similarity with previous models sequential
consistency is assumed.

We will now give the definition of the data dependencies implied by
location consistency assuming synchronization operations are
sequentially consistent.  The definition will not include control
dependencies implied by the mutual exclusion paradigm.  The definition
will be equivalent to location consistency for programs that conform
to the mutual exclusion paradigm, and it will extend location
consistency for programs that do not conform to the mutual exclusion
paradigm.  In the original definition of location consistency, due to
the mutual exclusion requirement there is an alternating order on the
synchronization operations to each variable: acquire, release,
acquire, release, etc.  Each acquire is immediately after a release
which is called its $most\_recent\_release$, and immediately before a
release by the same process.  The state of a variable, $x$, is defined
to be a partial order, $\prec$ which is the union of
$<_{PO}|(s,*,x,*)\cup(w,*,x,*)$ and the condition that all acquires to
$x$ are ordered after their $most\_recent\_release$.

Because $\prec$ is a partial order there may be many writes that could
be considered ``most recent'' in that there is no other write ordered
after them.  A read is allowed to return a value written by any one of
these most recent writes.  More formally, $\prec$ is augmented with
any process order edges between the read and any operation in
$(s,*,x,*)\cup(w,*,x,*)$ to produce $\prec'$.  Then, the read, $r$,
may return the value of any write, $w$, to the variable $x$ such that
$\not\exists w'$ where $w\prec' w'\prec' x$.  To put this in a similar
notation as the other synchronized models, the first requirement is
the same that synchronization operations must be sequentially
consistent.

\begin{quote}

$\exists<_{seq}=$SerialView$(<_{PO}|(s,*,*,*))$, and\\ $<_S=$the
transitive closure of $<_D\cup<_{seq}$ where $<_D$ is defined the same
as entry consistency

\end{quote}

For programs that obey mutual exclusion there is already a total
order, $<_{seq}$, on the synchronization operations to each variable.
So $<_S$ is merely the transitive closure of process order and
$most\_recent\_release$ order.  Therefore, $<_S$ is an equivalent
definition of $\prec$.  For programs that do not obey mutual exclusion
This is a sensible extension of the definition of $\prec$ that
maintains similarity with other synchronized models.  Now, location
consistency defines the set of values that may be returned by any
read.  To capture this, we will add to my formalism the notion of a
partial-ordered view.

\begin{definition}

\label{serialpartialview}

There exists a \emph{serial partial view} on a set of operations,
$subset$, respecting a partial order, $<$, denoted
SerialPartialView$(<|subset)$ iff

\begin{quote}

$\forall_{w\mapsto r\in O}\not\exists w'$ such that $w<w'<r$

\end{quote}

\end{definition}

A serial partial view is a minimal order, that is it doesn't add any
edges to $<$, it just checks if each read reads from a non dominated
write.  This is unlike a serial view that must add edges to create a
total order out of any partial order it respects.  The order, $<$,
must still be a partial order.  For example, there cannot exist a
serial partial view respecting a cyclic relation.  Now, location
consistency was defined where each read had its own serial partial
view.  However, if a serial partial view exists separately for two
reads over the same set of writes and synchronization operations, then
those two reads can be added to the same partial order, and it will
still be a serial partial view.  There is no interaction between the
two reads.  Therefore, the condition that all reads read a permissible
value can be stated thusly.

\begin{quote}

$\forall_{i\in P, x\in V}\exists$
SerialPartialView$(<_S\cup<_{PO}|(*,i,x,*)\cup(w,*,x,*))$

\end{quote}

Therefore, the definition of location consistency is identical to to
the definition of entry consistency with SerialView replaced by
SerialPartialView.

\section{Consistency Properties}

\label{lattice}

Some existing consistency models have been viewed as a combination of
other models.  For example, processor
consistency~\cite{goodman89:processor} is the combination of PRAM and
Cache consistency.  Causal ordering~\cite{ahamad91:causal} is the
transitive closure of process order and writes-to order.  Lamport's
original definition of sequential
consistency~\cite{lamport79:sequential} included a pair of properties
which, if independently enforced, would enforce sequential
consistency.  This suggests that perhaps many existing consistency
models could be viewed as different combinations of a few primitive
consistency properties.  In this section we define four such
properties.  Global Process Order (GPO) is the condition that there is
global agreement on the order of operations from each process.  Global
Data Order (GDO) is the condition that there is global agreement on
the order of operations to each variable.  Global Write-read-write
Order (GWO) is the condition that there is global agreement on the
order of potentially causally related writes.  Global Anti Order (GAO)
is the condition that there is global agreement on the order of any
two writes when a process can prove it has read one before the other.
Any combination of these properties results in a consistency model.
Enumerating these models results in the lattice shown in
Figure~\ref{fig-lattice}.

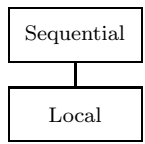
\begin{figure}[tp]

\begin{center}

\begin{picture}(50,50)

\put(0,30){\framebox(50,20){Sequential}}
\put(25,30){\line(0,-1){10}}
\put(0,0){\framebox(50,20){Local}}

\end{picture}

\end{center}

\caption{An initial consistency model lattice}

\label{lattice1}

\end{figure}

For pedagogical purposes, we will start with the lattice shown in
Figure~\ref{lattice1} and expand the lattice as properties are
developed.  The top of the lattice is defined to be sequential
consistency, and the bottom is defined to be local consistency as
these are the strongest and weakest properties in the literature (see
Section~\ref{related}.)

\subsection{Processor Consistency as a Combination of Properties}

Processor Consistency is defined to be a combination of PRAM and cache
consistency (see Definition~\ref{def-processor}.)  The given
definition of processor consistency requires constructing per-variable
views to satisfy cache consistency in addition to per-process views to
satisfy PRAM consistency.  To remove this inconvenience, we will
define two properties, one equivalent to PRAM consistency, and one
equivalent to cache consistency such that both properties can be
combined in the same per-process views.

\begin{definition}

An execution is \emph{Global Process Order (GPO)} iff

\begin{quote}

$\forall_{i\in P}\exists$
SerialView$(<_{iLocal}\cup<_{PO}|(*,i,*,*)\cup(w,*,*,*))$

\end{quote}

\end{definition}

\begin{theorem}

GPO is equivalent to PRAM consistency.

\end{theorem}

\begin{quotation}

Proof: The definitions of GPO and PRAM are identical.  The views for
GPO are required to respect local order for similarity with the
properties to follow.  However, this requirement is redundant because
process order is a superset of local order for any process.

\end{quotation}

\begin{definition}

Two operations are ordered by \emph{data order}, $o_1<_{DO}o_2$, iff
they are to the same variable, and either

\begin{enumerate}

\item $o_1<_{PO}o_2$, or

\item $o_1\mapsto o_2$, or

\item \label{pre-read} There exists a read, $r$, to the same variable
such that $o_1<_{PO}r$, $o_1$ has a different value than $r$, and
$o_2\mapsto r$, or

\item There exists an operation, $o$, such that $o_1<_{DO}o<_{DO}o_2$.

\end{enumerate}

\end{definition}

Data order captures the restrictions involved in constructing the
required views for cache consistency.  The operations $o_1$ and $o_2$
can be either reads or writes, but must be to the same variable.  Data
order contains writes-to order and process order restricted to pairs
of operations to the same variable because the views for cache
consistency must be serial and respect process order.  For the third
condition, a particular process reads or writes a value, $o_1$, and
then at a later time reads a different value from the same variable,
$r$.  That process can deduce that a write, $o_2$, must have occurred
between those two operations and so the restriction is included in
data order.  The fourth condition requires that data order is a
transitive closure.

\begin{definition}

An execution is \emph{Global Data Order (GDO)} iff

\begin{quote}

$\forall_{i\in P}\exists$
SerialView$(<_{iLocal}\cup<_{DO}|(*,i,*,*)\cup(w,*,*,*))$

\end{quote}

\end{definition}

The proof that GDO is equivalent to cache consistency uses several
lemmas:

\begin{lemma}

\label{GDO1}

If an execution is Cache Consistent then Data Order is acyclic.

\end{lemma}

\begin{quotation}

Proof: Data order only contains edges between pairs of operations to
the same variable.  Therefore, if data order were cyclic, the cycle
would have to involve only operations to a single variable.  Cache
consistency requires for every variable a serial view respecting
process order on all the operations to that variable. We will show
that these views must also respect data order, and so data order is
acyclic.

The cache consistent view for a variable respects process order by
definition and writes-to order because it is serial so it respects the
first two conditions of data order.  The third condition of data order
must also be respected.  If $o_1$ is process ordered before $r$ it
must come before $r$ in the view.  If, in addition, $o_1$ has a
different value than $r$, and $o_2$ writes to $r$, then $o_1$ must
come before $o_2$ in the view.  If not and $o_1$ is a write then $r$
does not read from the most recent write so the view is not serial.
If not and $o_1$ is a read then either $o_2$ is the most recent write
before $o_1$, in which case $o_1$ does not read from the most recent
write, or there is another write between $o_2$ and $o_1$.  This write
is also between $o_2$ and $r$, and it has the same value as $o_1$
which is different than $r$, so $r$ does not read from the most recent
write and the view is not serial.  Since the view is a total order and
it respects the first three conditions of data order it must respect
their transitive closure which is the fourth condition of data order.

The views required for cache consistency must respect data order.  If
data order contained a cycle then the view for some variable could not
be constructed and the execution would not be cache consistent.
Therefore, if the execution is cache consistent data order is
acyclic.

\end{quotation}

\begin{lemma}

\label{do-reads}

If two reads are ordered by data order then either they are by the
same process, or they are ordered by a transitive chain containing a
write.

\end{lemma}

\begin{quotation}

Proof: Two reads cannot be ordered by writes-to, or by having one
write to a read that the other is process ordered before.  So the only
way two reads can be data ordered is by process order, or a transitive
chain.  If two reads are not by the same process, and are data ordered
take the last operation in the transitive chain.  If this operation is
a write it satisfies the lemma.  Otherwise, it must be a read by the
same process as the final read.  By the same logic the next to last
operation in the chain must also be a write, or a read by the same
process as the final read.  By induction, if there is no write in the
chain then the first operation in the chain must be a read by the same
process as the final read.  Therefore, if the two reads are not by the
same process there must be a write in the transitive chain.

\end{quotation}

\begin{lemma}

\label{GDO2}

If an execution is GDO then data order is acyclic.

\end{lemma}

\begin{quotation}

Proof: GDO requires a view for every process that is serial and
respects data order over the subset of all operations by that process
plus all writes.  If these views are constructible then data order
must be acyclic at least on the subsets of operations in each view.
Therefore, if data order is cyclic then the cycle must contain at
least two read operations by different processes, $r_1$ and $r_2$,
such that $r_1<_{DO}r_2$ and $r_2<_{DO}r_1$.  By Lemma~\ref{do-reads}
these two reads must be ordered by two transitive chains, and each
chain must contain a write.  Because data order is a transitive
closure there must be a cycle between the writes in the two chains.
This makes it impossible to construct the views required for GDO
because every view must include all writes.  If data order is cyclic
then the views required for GDO cannot be constructed.  Therefore, if
an execution is GDO then data order is acyclic.

\end{quotation}

\begin{lemma}

\label{GDO3}

If data order is acyclic then

\begin{quote}

$\forall_{x\in V}\exists<_x=$SerialView$(<_{DO}|(*,*,x,*))$

\end{quote}

\end{lemma}

\begin{quotation}

Proof: First, collect all the operations on a single variable and
place them into groups where each group contains a write and all reads
that the write writes-to.  Order the operations in each group in an
order that respects data order.  This is possible because data order
is acyclic.  The reads in a group all read from the write in that
group so the write will be ordered first in each group.  The serial
view for that variable is constructed by ordering the groups with no
interleaving of operations between different groups.  For every read,
the most recent write to the same variable must be the write from its
group which is the write which wrote-to it so the view must be serial.
Any order of the groups with no interleaving will produce a serial
view.

If $G_1$ and $G_2$ are two groups then define group order, $<_{GO}$
as: $G_1<_{GO}G_2$ iff $\exists o_1\in G_1,o_2\in G_2$ such that
$o_1<_{DO}o_2$.  If group order is acyclic then any topological sort
on group order will produce a view that respects data order and is
serial.  Assume there is a cycle in group order, but not in data
order.  Take any two ordered groups from the cycle, $G_1<_{GO}G_2$.
We will show that the writes from the groups, $w_1$ and $w_2$, must be
ordered by data order.  Therefore, any cycle in group order must be
accompanied by a cycle in data order.  So if data order is acyclic
then group order must be acyclic and the views can be constructed.

There must be operations from the two groups such that $o_1<_{DO}o_2$.
Either $o_1$ is $w_1$, or $o_1$ is a read that $w_1$ writes-to.  So
$w_1<_{DO}o_2$.  Also, either $o_2$ is $w_2$ in which case
$w_1<_{DO}w_2$, or $o_2$ is a read that $w_2$ writes-to.  If $o_2$ is
a read consider how it came about that $w_1$ is data ordered before
$o_2$.  $W_1$ did not write to $o_2$ so either $w_1<_{PO}o_2$, or they
are ordered by a transitive chain.  If $w_1<_{PO}o_2$ then
$w_1<_{DO}w_2$ because $w_2\mapsto o_2$.  If not, let $o$ be the last
operation in the transitive chain so $w_1<_{DO}o<_{DO}o_2$.  Either
$o\mapsto o_2$ in which case $o$ is $w_2$ and $w_1<_{DO}w_2$ or
$o<_{PO}o_2$ in which case $o<_{DO}w_2$ because $w_2\mapsto o_2$ so
$w_1<_{DO}w_2$.

Therefore, if $G_1<_{GO}G_2$ then $w_1<_{DO}w_2$.  Any cycle in group
order will be reflected in data order by the writes.  If data order is
acyclic then there can be no cycle in group order, and a topological
sort of the groups respecting group order will produce the required
serial view respecting data order for that variable.

\end{quotation}

\begin{lemma}

\label{GDO-iLocal}

If the serial views, $<_x$, defined in Lemma~\ref{GDO3} exist then

\begin{quote}

$\forall_{i\in P}\cup_{x\in V}<_x\bigcup<_{iLocal}$ is acyclic.

\end{quote}

\end{lemma}

\begin{quotation}

Proof: Assume that a cycle exists for some process, $i$.  The views,
$<_x$, are acyclic, and their union cannot contain a cycle because no
operation is in more than one view.  Therefore, the cycle must have an
edge in the local order, and thus include operations by process $i$.
Pick any operation by process $i$ in the cycle.  Call it $o$.  Follow
the edges that make up the cycle.  If you follow an edge in local
order then you must reach an operation by process $i$ that occurs
after $o$ in local order.  If you follow an edge not in local order
then you must reach an operation to the same variable as $o$ by a
process other than $i$.  Operations by other processes are not ordered
by local order, and thus the cycle must proceed through operations to
the same variable following edges of $<_x$ for that variable until
reaching an operation by process $i$.  This operation must be to the
same variable as $o$, and must be ordered after $o$ by local order.
Otherwise, the view for that variable would not respect data order.

In any case, the first operation by process $i$ encountered after $o$
in the cycle must be after $o$ in local order.  Call this operation
$o'$.  By the same logic the next operation by process $i$ after $o'$
in the cycle must be ordered after $o'$ and $o$ by local order.  By
induction, every operation by process $i$ in the cycle must be after
$o$ in local order.  Eventually the cycle will reach $o$ itself
showing that there is a cycle in local order which is a
contradiction.

\end{quotation}

\begin{lemma}

\label{GDO4}

If the views, $<_x$, defined in Lemma~\ref{GDO3} exist then the
execution is GDO.

\end{lemma}

\begin{quotation}

Proof: Construct the view for process $i$ required for GDO as any
topological sort of $(*,i,*,*)\cup(w,*,*,*)$ respecting $\cup_{x\in
V}<_x\bigcup<_{iLocal}$ This is possible because by
Lemma~\ref{GDO-iLocal} the relation is acyclic.  The views will be
serial because the views, $<_x$, are serial, and the relative position
of all pairs of operations to the same variable is preserved.  Data
order only contains edges between two operations to the same variable
and so is a subset of $\cup_{x\in V}<_x$.  Therefore, the constructed
views respect local order and data order, and they are serial so the
execution is GDO.

\end{quotation}

\begin{lemma}

\label{GDO5}

If the views, $<_x$, defined in Lemma~\ref{GDO3} exist then the
execution is cache consistent.

\end{lemma}

\begin{quotation}

Proof: Process order restricted to the set of operations on a single
variable is a subset of data order.  Therefore, any view on
$(*,*,x,*)$ that respects data order will also respect process order.
Therefore, the views defined in Lemma~\ref{GDO3} respect process
order, and so prove that the execution is cache consistent.

\end{quotation}

\begin{theorem}

\label{acyclic}

An execution is Cache Consistent iff it is GDO iff data order is
acyclic.

\end{theorem}

\begin{quotation}

Proof: Follows directly from
lemmas~\ref{GDO1},~\ref{GDO2},~\ref{GDO3},~\ref{GDO4},~and~\ref{GDO5}.

\end{quotation}

This is an important result because it provides two new ways to define
cache consistency.  One can determine whether an execution is cache
consistent by the original method of constructing per-variable serial
views, or now by constructing per-process serial views, or even by
testing the cyclicity of the data order relation.  Now that cache
consistency is defined over per-process views we can combine GPO and
GDO more easily.

\begin{definition}

An execution is \emph{GPO+GDO} iff

\begin{quote}

$\forall_{i\in P}\exists$
SerialView$(<_{iLocal}\cup<_{PO}\cup<_{DO}|(*,i,*,*)\cup(w,*,*,*))$

\end{quote}

\end{definition}

However, GPO+GDO is not quite equivalent to Goodman's definition of
processor consistency.  Processor consistency requires that all
processes agree on a total order of operations to each variable.  In
Figure~\ref{GPOplusGDO}, the processes cannot agree on the order of
the writes to $z$.  If $(w,p_1,z,2)$ was first, then $p_2$ should have
read $1$ from $x$.  Likewise, if $(w,p_2,z,4)$ was first, then $p_1$
should have read $3$ from $y$.  However, the two writes to $z$ are not
ordered by data order.  Even under processor consistency they are
allowed to occur in either order, but GPO+GDO does not enforce that
they be seen in the same order by all processes.  This can be solved
by creating augmented data order, $<_{DO'}$.  Augmented data order is
any superset of data order that enforces a total order on all
operations to each variable.  By Theorem~\ref{acyclic}, any GDO
execution respects at least one augmented data order.  The problem is
that there may be more than one, and a single augmented data order may
not be consistent with process order at all sites.  GPO+GDO' is
defined similarly to GPO+GDO.

\begin{figure}[tp]

\begin{center}

\begin{tabular}{l}

$(w,p_1,x,1)<_{PO}(w,p_1,z,2)<_{PO}(r,p_1,y,\bot)$ \\
$(w,p_2,y,3)<_{PO}(w,p_2,z,4)<_{PO}(r,p_2,x,\bot)$ \\
$(w,\epsilon,x,\bot)\mapsto(r,p_2,x,\bot)$         \\
$(w,\epsilon,y,\bot)\mapsto(r,p_1,y,\bot)$

\end{tabular}

\end{center}

\caption{A GPO+GDO, but not processor consistent execution.}

\label{GPOplusGDO}

\end{figure}

\begin{theorem}

Goodman's definition of processor consistency (as given in
Subsection~\ref{non-synch}) is equivalent to GPO+GDO'.

\end{theorem}

\begin{quotation}

Proof: Augmented data order is equivalent to the per-variable cache
consistency views required for processor consistency.  The per-process
views for GPO+GDO' respect process order and augmented data order.
The per-process views for processor consistency respect process order,
and the per-variable cache consistent views.  Therefore, the two
required sets of views are equivalent.

\end{quotation}

Augmented data order solves the problem of equivalence to Goodman's
definition of processor consistency.  However, we feel that even
without augmented data order GPO+GDO is in line with the intended
purpose of process order.  In Figure~\ref{GPOplusGDO} the writes to
$z$ are unordered.  Inserting reads to $z$ to detect the order of the
writes would create data order dependencies and eliminate the need for
augmented data order.  Is it likely that the correctness of a program
would depend on the fact that those two operations are seen in the
same order by all processes when their order is unknown?  Also, the
execution in Figure~\ref{GPOplusGDO} was taken
from~\cite{ahamad92:processor} as an example of an execution accepted
by the DASH definition of processor consistency, and rejected by
Goodman's definition.  The space of consistency models surrounding
processor consistency has not been completely searched.  We believe
that GPO+GDO will prove to be a useful consistency model, and a
systematic examination of this search space will lead to greater
understanding of the foundations of consistency models.

Alternatively, GPO+GDO is equivalent to the following modified
definition of processor consistency where the $\forall_{i\in P}$ is
moved outside of the $\forall_{x\in V}$, and each process respects a
set of cache consistent views, but all processes do not have to
respect the same set of views.

\begin{quote}

$\forall_{i\in P}\forall_{x\in
V}\exists<_x=$SerialView$(<_{PO}|(*,*,x,*))$, and\\ $\exists$
SerialView$((\cup_{x\in V}<_x)\bigcup<_{PO}|(*,i,*,*)\cup(w,*,*,*))$

\end{quote}

This issue is discussed in more detail in Section~\ref{transitions}.
The same issue comes up when defining synchronized consistency models
as consistency transitions.  The synchronization operations must be
sequentially consistent, but there may be more than one total order
that would satisfy sequentially consistency.  The ordinary operations
are not required to be sequentially consistent, and may demonstrate
that different processes saw different sequential orders even though
the synchronization operations in isolation are sequentially
consistent.

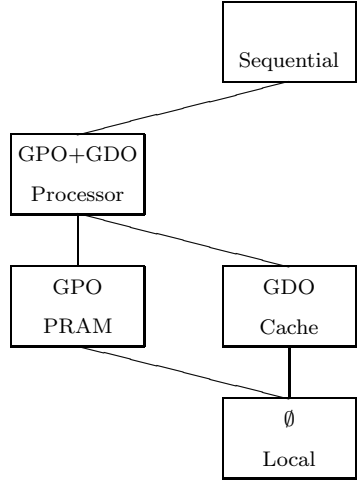
\begin{figure}[tp]

\begin{center}

\begin{picture}(210,180)

\put(80,150){\framebox(50,30){}}
\put(80,150){\makebox(50,15){Sequential}}

\put(105,150){\line(-4,-1){80}}

\put(0,100){\framebox(50,30){}}
\put(0,115){\makebox(50,15){GPO+GDO}}
\put(0,100){\makebox(50,15){Processor}}

\put(25,100){\line(0,-1){20}}
\put(25,100){\line(4,-1){80}}

\put(0,50){\framebox(50,30){}}
\put(0,65){\makebox(50,15){GPO}}
\put(0,50){\makebox(50,15){PRAM}}
\put(80,50){\framebox(50,30){}}
\put(80,65){\makebox(50,15){GDO}}
\put(80,50){\makebox(50,15){Cache}}

\put(25,50){\line(4,-1){80}}
\put(105,50){\line(0,-1){20}}

\put(80,0){\framebox(50,30){}}
\put(80,15){\makebox(50,15){$\emptyset$}}
\put(80,0){\makebox(50,15){Local}}

\end{picture}

\end{center}

\caption{A consistency model lattice including processor consistency}

\label{lattice2}

\end{figure}

GPO+GDO begins a framework for defining consistency properties (see
Figure~\ref{lattice2}.)  Any property that can be defined as a
relation which must be respected by per-process views can be combined
with process order and data order to create new consistency models.

\begin{figure}[tp]

\begin{center}

\begin{tabular}{l}

$(w,p_1,x,1)<_{PO}(w,p_1,y,2)$  \\
$(r,p_2,y,2)<_{PO}(w,p_2,x,3)$  \\
$(r,p_3,x,3)<_{PO}(r,p_3,x,1)$  \\
$(w,p_1,x,1)\mapsto(r,p_3,x,1)$ \\
$(w,p_1,y,2)\mapsto(r,p_2,y,2)$ \\
$(w,p_2,x,3)\mapsto(r,p_3,x,3)$

\end{tabular}

\end{center}

\caption{A PRAM and cache, but not GPO+GDO consistent execution.}

\label{GPOcapGDO}

\end{figure}

There can also be executions that are GPO and GDO, but not GPO+GDO.
Ahamad, et. al.~\cite{ahamad92:processor} provide the execution in
Figure~\ref{GPOcapGDO} which is PRAM and cache consistent, but not
processor consistent.  The execution is GPO because of the following
views.

\begin{quote}

$p_1:(w,p_2,x,3)<_{p_1}(w,p_1,x,1)<_{p_1}(w,p_1,y,2)$\\
$p_2:(w,p_1,x,1)<_{p_2}(w,p_1,y,2)<_{p_2}(r,p_2,y,2)<_{p_2}(w,p_2,x,3)$\\
$p_3:(w,p_2,x,3)<_{p_3}(r,p_3,x,3)<_{p_3}(w,p_1,x,1)<_{p_3}(r,p_3,x,1)<_{p_3}(w,p_1,y,2)$

\end{quote}

Data order is as follows.

\begin{quote}

$(w,p_2,x,3)<_{DO}(r,p_3,x,3)<_{DO}(w,p_1,x,1)<_{DO}(r,p_3,x,1)$ \\
$(w,p_1,y,2)<_{DO}(r,p_2,y,2)$

\end{quote}

The execution is GDO because of the following views.

\begin{quote}

$p_1:(w,p_2,x,3)<_{p_1}(w,p_1,x,1)<_{p_1}(w,p_1,y,2)$\\
$p_2:(w,p_1,y,2)<_{p_2}(r,p_2,y,2)<_{p_2}(w,p_2,x,3)<_{p_2}(w,p_1,x,1)$\\
$p_3:(w,p_2,x,3)<_{p_3}(r,p_3,x,3)<_{p_3}(w,p_1,x,1)<_{p_3}(r,p_3,x,1)<_{p_3}(w,p_1,y,2)$

\end{quote}

However, in $<_{p_2}$ the position of $(w,p_1,x,1)$ is different
between the GPO and GDO views.  There is no view $<_{p_2}$ that
conform to both $<_{PO}$ and $<_{DO}$.

\begin{quote}

$(w,p_1,x,1)<_{PO}(w,p_1,y,2)<_{DO}(r,p_2,y,2)<_{PO}(w,p_2,x,3)<_{DO}(w,p_1,x,1)$

\end{quote}

so $<_{DO}\cup<_{PO}$ has a cycle.  $<_{p_2}$ must contain all of
these operations and thus cannot be constructed.  This leads to the
definition of another consistency model.

\begin{definition}

An execution is \emph{GPO$\cap$GDO} iff

\begin{quote}

$\forall_{i\in P}\exists$
SerialView$(<_{iLocal}\cup<_{PO}|(*,i,*,*)\cup(w,*,*,*))\bigwedge$\\
$\forall_{i\in P}\exists$
SerialView$(<_{iLocal}\cup<_{DO}|(*,i,*,*)\cup(w,*,*,*))$

\end{quote}

\end{definition}

Any pair of properties can be combined in this way creating a new
consistency model.  The meaning of these models has not been explored
previously in the literature, and we have not explored them in our
work.  They are mentioned here for completeness.

\subsection{Causal Consistency as a Combination of Properties}

Causal consistency is stronger than GPO, but incomparable to GPO+GDO.
Therefore, there should be a property that enforces that part of
causal not already covered by GPO.  Causal consistency depends on the
causal relation which is the transitive closure of process order and
writes-to order.  The causal relation is made up of three types of
edges: edges in process order, edges in writes-to order, and edges not
in either order, but in the transitive closure.  Process order has
already been identified as a primitive property, and any serial view
respects writes-to order.  Therefore, we now define another property
which contains the edges in the transitive closure. This new property
can be used with process order to define causal consistency.

\begin{figure}[tp]

\begin{center}

\begin{tabular}{l}

$(w,p_1,x,1)<_{PO}(r,p_1,y,3)<_{PO}(r,p_1,x,1)$ \\
$(r,p_2,x,1)<_{PO}(w,p_2,x,2)<_{PO}(w,p_2,y,3)$ \\
$(w,p_1,x,1)\mapsto(r,p_1,x,1)$ \\
$(w,p_1,x,1)\mapsto(r,p_2,x,1)$ \\
$(w,p_2,y,3)\mapsto(r,p_1,y,3)$

\end{tabular}

\end{center}

\caption{An Execution That Violates Causal Consistency}

\label{not-causal}

\end{figure}

For example, Figure~\ref{not-causal} contains an execution that is not
causally consistent even though the following serial views respect
both process order and writes-to order.

\begin{quote}

$p_1:(w,p_2,x,2)<_{p_1}(w,p_2,y,3)<_{p_1}(w,p_1,x,1)<_{p_1}(r,p_1,y,3)<_{p_1}(r,p_1,x,1)$\\
$p_2:(w,p_1,x,1)<_{p_2}(r,p_2,x,1)<_{p_2}(w,p_2,x,2)<_{p_2}(w,p_2,y,3)$

\end{quote}

There is a causal dependency from $(w,p_1,x,1)$ to $(w,p_2,x,2)$
because

\begin{quote}

$(w,p_1,x,1)\mapsto(r,p_2,x,1)<_{PO}(w,p_2,x,2)$.

\end{quote}

However, $<_{p_1}$ places them in the opposite order because $<_{p_1}$
does not contain the operation $(r,p_2,x,1)$ which is a read operation
by $p_2$.  Therefore, it violates neither process order nor writes-to
order among the operations in its view.  To be causally consistent the
view for each process must respect:

\begin{quote}

(the transitive closure of $<_{PO}\cup\mapsto)|(*,i,*,*)\cup(w,*,*,*)$

\end{quote}

The definition for GPO already respects:

\begin{quote}

the transitive closure of $(<_{PO}\cup\mapsto|(*,i,*,*)\cup(w,*,*,*))$

\end{quote}

Note the different parentheses.  The new property can be found in the
set difference of these two relations.  For an edge to be in the first
relation and not the second, two operations in
$(*,i,*,*)\cup(w,*,*,*)$ must be transitively ordered by a chain of
operation not in $(*,i,*,*)\cup(w,*,*,*)$.  The only operations not in
that set are reads by a process other than $i$.  Reads cannot be
ordered with each other by writes-to order, and if a chain of reads is
ordered by process order they must all be by the same process, and the
first and last reads in the chain will be ordered.  So, any transitive
chains of the sort we are interested in must have an operation, $o_1$
ordered by process order or writes-to order before a read, $r_1$,
possibly ordered by process order before another read, $r_2$, ordered
by process order or writes-to order before an operation, $o_2$.  All
possibilities are summarized in Figure~\ref{enumerated-transitive}:

\begin{figure}[tp]

\begin{center}

\begin{tabular}{llllllll}

1.&$o_1$&$\mapsto$&$r_1$&$<_{PO}$&$r_2$&$\mapsto$&$o_2$\\
2.&$o_1$&$\mapsto$&$r_1$&        &     &$\mapsto$&$o_2$\\
3.&$o_1$&$<_{PO}$ &$r_1$&$<_{PO}$&$r_2$&$\mapsto$&$o_2$\\
4.&$o_1$&$<_{PO}$ &$r_1$&        &     &$\mapsto$&$o_2$\\
5.&$o_1$&$<_{PO}$ &$r_1$&$<_{PO}$&$r_2$&$<_{PO}$ &$o_2$\\
6.&$o_1$&$<_{PO}$ &$r_1$&        &     &$<_{PO}$ &$o_2$\\
7.&$o_1$&$\mapsto$&$r_1$&$<_{PO}$&$r_2$&$<_{PO}$ &$o_2$\\
8.&$o_1$&$\mapsto$&$r_1$&        &     &$<_{PO}$ &$o_2$

\end{tabular}

\end{center}

\caption{Enumerated Possibilities for a Causal Transitive Chain}

\label{enumerated-transitive}

\end{figure}

Cases~1,~2,~3, and~4 are impossible because a read cannot be on the
left hand side of a writes-to relation.  In cases~5 and~6, the two
operations, $o_1$ and $o_2$, are already ordered by process order.  In
case~7, $r_1$ and $o_2$ are ordered by process order so it reduces to
case 8.  Therefore, the only case that must be considered is case~8.

In case~8, $o_1$ must be a write because it writes to $r_1$.  $o_2$ is
in the set $(*,i,*,*)\cup(w,*,*,*)$.  $r_1$ is not in this set and so
is not by process $i$.  $o_2$ is by the same process as $r_1$ so it
must be a write by another process.  Therefore, only causal chains
between two writes must be considered.

\begin{definition}

Two writes are ordered by \emph{write-read-write order},
$w_1<_{WO}w_2$, iff there exists a read, $r$ such that $w_1\mapsto
r<_{PO}w_2$.

\end{definition}

\begin{definition}

An execution is \emph{Global Write-read-write Order (GWO)} iff

\begin{quote}

$\forall_{i\in P}\exists$
SerialView$(<_{iLocal}\cup<_{WO}|(*,i,*,*)\cup(w,*,*,*))$

\end{quote}

\end{definition}

\begin{theorem}

GPO+GWO is equivalent to causal consistency.

\end{theorem}

\begin{quotation}

Proof: By the logic above, the transitive closure of
$<_{PO}\cup<_{WO}\cup\mapsto|(*,i,*,*)\cup(w,*,*,*)$ is equivalent to
$<_{CR}|(*,i,*,*)\cup(w,*,*,*)$.  Any serial view respects $\mapsto$,
and a view is a total order so if it respects a relation it respects
the transitive closure of that relation.  Also, any view that respects
$<_{PO}$ respects $<_{iLocal}$.  So a serial view respects
$<_{iLocal}\cup<_{PO}\cup<_{WO}|(*,i,*,*)\cup(w,*,*,*)$ iff it
respects $<_{CR}|(*,i,*,*)\cup(w,*,*,*)$.  The first is the
requirement for GPO+GWO.  The second is the requirement for causal
consistency.

\end{quotation}

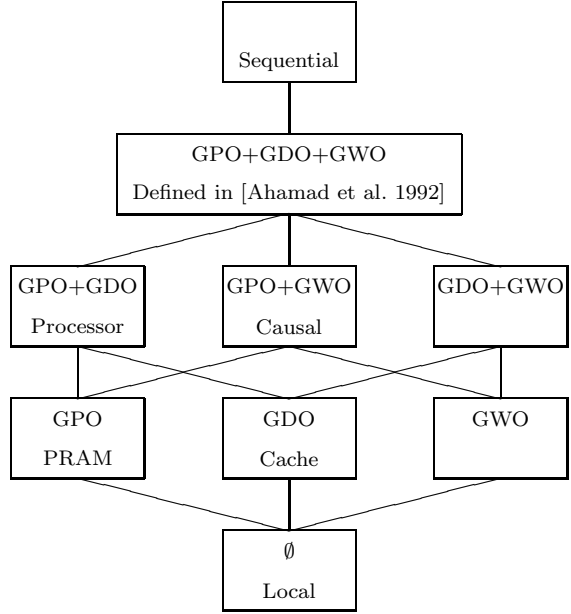
\begin{figure}[tp]

\begin{center}

\begin{picture}(210,230)

\put(80,200){\framebox(50,30){}}
\put(80,200){\makebox(50,15){Sequential}}

\put(105,200){\line(0,-1){20}}

\put(40,150){\framebox(130,30){}}
\put(40,165){\makebox(130,15){GPO+GDO+GWO}}
\put(40,150){\makebox(130,15){Defined in~\cite{ahamad92:processor}}}

\put(105,150){\line(-4,-1){80}}
\put(105,150){\line(0,-1){20}}
\put(105,150){\line(4,-1){80}}

\put(0,100){\framebox(50,30){}}
\put(0,115){\makebox(50,15){GPO+GDO}}
\put(0,100){\makebox(50,15){Processor}}
\put(80,100){\framebox(50,30){}}
\put(80,115){\makebox(50,15){GPO+GWO}}
\put(80,100){\makebox(50,15){Causal}}
\put(160,100){\framebox(50,30){}}
\put(160,115){\makebox(50,15){GDO+GWO}}

\put(25,100){\line(0,-1){20}}
\put(25,100){\line(4,-1){80}}
\put(105,100){\line(-4,-1){80}}
\put(105,100){\line(4,-1){80}}
\put(185,100){\line(-4,-1){80}}
\put(185,100){\line(0,-1){20}}

\put(0,50){\framebox(50,30){}}
\put(0,65){\makebox(50,15){GPO}}
\put(0,50){\makebox(50,15){PRAM}}
\put(80,50){\framebox(50,30){}}
\put(80,65){\makebox(50,15){GDO}}
\put(80,50){\makebox(50,15){Cache}}
\put(160,50){\framebox(50,30){}}
\put(160,65){\makebox(50,15){GWO}}

\put(25,50){\line(4,-1){80}}
\put(105,50){\line(0,-1){20}}
\put(185,50){\line(-4,-1){80}}

\put(80,0){\framebox(50,30){}}
\put(80,15){\makebox(50,15){$\emptyset$}}
\put(80,0){\makebox(50,15){Local}}

\end{picture}

\end{center}

\caption{A consistency model lattice including causal consistency}

\label{lattice3}

\end{figure}

Adding GWO to the evolving lattice of consistency models results in
Figure~\ref{lattice3}.  The model GPO+GDO+GWO has been previously
discovered.  In~\cite{ahamad92:processor} the authors noticed that the
definition of processor consistency allows executions that violate
causality, and they developed an extension to processor consistency to
prevent this.  At this point the lattice contains two new consistency
models: GWO, and GDO+GWO.

\subsection{Sequential Consistency as a Combination of Properties}

\label{anti-property}

\begin{figure}[tp]

\begin{center}

\begin{tabular}{l}

$(w,p_1,x,1)<_{PO}(r,p_1,y,\bot)<_{PO}(r,p_1,y,2)$\\
$(w,p_2,y,2)<_{PO}(r,p_2,x,\bot)<_{PO}(r,p_2,x,1)$\\
$(w,\epsilon,y,\bot)\mapsto(r,p_1,y,\bot)$\\
$(w,p_2,y,2)\mapsto(r,p_1,y,2)$\\
$(w,\epsilon,x,\bot)\mapsto(r,p_2,x,\bot)$\\
$(w,p_1,x,1)\mapsto(r,p_2,x,1)$

\end{tabular}

\end{center}

\caption{An Execution that Violates Sequential Consistency.}

\label{non-sequential}

\end{figure}

GPO+GDO+GWO is weaker than sequential consistency.  Consider the
execution in Figure~\ref{non-sequential}.  The two writes are not by
the same processor, nor to the same variable, and they are not
causally related.  These two writes could be seen to occur in either
order, but to be sequentially consistent every process must see them
in the same order.  In this execution the following cycle exists:

\begin{quote}

$(w,p_1,x,1)<_{PO}(r,p_1,y,\perp)<_{DO}(w,p_2,y,2)<_{PO}(r,p_2,x,\perp)<_{DO}(w,p_1,x,1)$

\end{quote}

But GPO+GDO+GWO requires separate views for processes $p_1$ and $p_2$,
and each view includes only its own read operations.  So the following
views are acceptable:

\begin{quote}

$p_1:(w,p_1,x,1)<_{p_1}(r,p_1,y,\perp)<_{p_1}(w,p_2,y,2)<_{p_1}(r,p_1,y,2)$\\
$p_2:(w,p_2,y,2)<_{p_2}(r,p_2,x,\perp)<_{p_2}(w,p_1,x,1)<_{p_2}(r,p_2,x,1)$

\end{quote}

For this execution to be prohibited there must be another order that
takes a cycle which includes read operations and creates a cycle among
only write operations.  In Figure~\ref{enumerated-transitive} there
were eight cases of a causal transitive chain.  Four of them were
deemed impossible because a read could not be on the left hand side of
a writes-to order.  These cases are made possible by using data order
as a generalization of writes-to order.  A read may not be able to
write to another operation, but it may be able to prove that it
happened first.  These four cases are the basis of a new consistency
property called \emph{anti order}.  The name anti order comes from
parallel compiler optimization.  When a program contains a read and
later a write to the same variable their orders cannot be reversed.
This is called an anti dependency, and is similar to this situation
where a read can prove, through data order, that a write happened
after it.  It is at least similar enough to borrow the name.

The purpose of Global Anti Order (GAO) is to complete the set of
consistency properties so that, together, they simulate sequential
consistency.  To do this, anti order must take cycles involving read
operations, and short circuit them to produce cycles involving only
write operations.  Therefore, anti order is limited to the case where
$o_1$ and $o_2$ (in Figure~\ref{enumerated-transitive}) are writes.
This weakens anti order, and our desire is to produce the weakest
relation that supports the assertion that GPO+GDO+GWO+GAO is
equivalent to sequential consistency.  From
Figure~\ref{enumerated-transitive}, case 1 seems necessary because the
writes may only be ordered through the reads.  Case 2 seems
unnecessary because the writes are already ordered by data order, but
it will be needed as explained later.  Case 3 reduces to case 4, and
case 4 solves the problem of Figure~\ref{non-sequential} since

\begin{quote}

$(w,p_1,x,1)<_{PO}(r,p_1,y,\perp)<_{DO}(w,p_2,y,2)$, and\\
$(w,p_2,y,2)<_{PO}(r,p_2,x,\perp)<_{DO}(w,p_1,x,1)$, so\\
$(w,p_1,x,1)<_{AO}(w,p_2,y,2)<_{AO}(w,p_1,x,1)$

\end{quote}

So, an initial idea is to base anti order on only cases 1 and 4.
However, this solution is not complete.  The execution in
Figure~\ref{non-sequential} can be modified by removing the final read
of each process.  This means that condition~\ref{pre-read} of data
order no longer applies and the writes are not data ordered after
$(r,p_1,y,\perp)$ and $(r,p_2,x,\perp)$.  There is no anti order
cycle, and the execution is no longer rejected by anti order even
though it still violates sequential consistency.  The problem is with
a limitation of data order.  If a read does not read from a write to
the same variable this is not enough to deduce that the write happened
after the read.  It could have happened very early and been
overwritten.  However, it could not have happened between the read and
the write that wrote-to the read.  This ordering restriction is not
present in data order.  Capturing this restriction requires a
non-deterministic order called \emph{serial order}.  One can think of
serial order as ``pseudo data order'' that can replace writes-to order
in the cases given in Figure~\ref{enumerated-transitive}.  We now need
to include case 2 because $w_1\mapsto r_1<_{SO}w_2$ does not guarantee
that the writes are data ordered.

\begin{definition}

A \emph{Serial Order}, $<_{SO}$, is a minimal set of edges that
enforces the following condition:

\begin{quote}

$\forall_{w,r\in O}$ such that $w$ and $r$ are to the same variable
and do not have the same value either $w<_{SO}w'\mapsto r$ or
$r<_{SO}w$

\end{quote}

\end{definition}

So the final definition of anti order is as follows.

\begin{definition}

\emph{Anti-Order}, $<_{AO(<_{SO})}$,

\begin{quote}

\begin{tabbing}
1234\=1234\=\kill

Given a serial order, $<_{SO}$,\\
$\forall_{w_1,w_2\in O}\ w_1<_{AO}w_2$ iff\\
\>$\exists r_1,r_2$ such that\\
\>\>$w_1\mapsto r_1<_{PO}r_2<_{DO}w_2$, or\\
\>\>$w_1\mapsto r_1<_{PO}r_2<_{SO}w_2$, or\\
\>\>$w_1\mapsto r_1<_{SO}w_2$, or\\
\>\>$w_1<_{PO}r_1<_{DO}w_2$, or\\
\>\>$w_1<_{PO}r_1<_{SO}w_2$

\end{tabbing}

\end{quote}

\end{definition}

To define global anti order there must be serial views that respect
anti order for some definition of serial order.  However, this is
still not enough.  In the example of Figure~\ref{non-sequential} with
the final reads removed serial order could be defined as:

\begin{quote}

$(w,p_1,x,1)<_{SO}(w,\epsilon,x,\perp)$\\
$(w,p_2,y,2)<_{SO}(w,\epsilon,y,\perp)$

\end{quote}

There would be no anti order links.  The views could then be written:

\begin{quote}

$p_1:(w,p_1,x,1)<_{p_1}(r,p_1,y,\perp)<_{p_1}(w,p_2,y,2)$\\
$p_2:(w,p_2,y,2)<_{p_2}(r,p_2,x,\perp)<_{p_2}(w,p_1,x,1)$

\end{quote}

These views respect process order, data order, write-read-write order,
and even anti order for some definition of serial order.  They also
respect some definition of serial order, but not the same definition
that was used to construct anti order.  This is the crucial last piece
of the puzzle.  The views must respect the same definition of serial
order that was used to construct anti order.

\begin{definition}

An execution is \emph{Global Anti Order (GAO)} iff $\exists<_{SO}$
such that

\begin{quote}

$\forall_{i\in P}\exists$
SerialView$(<_{iLocal}\cup<_{SO}\cup<_{AO(<_{SO})}|(*,i,*,*)\cup(w,*,*,*))$

\end{quote}

\end{definition}

Serial order is a non-deterministic order in the sense that it may
have many possible definitions, and if any one of the definitions
accepts the execution then the execution is accepted.  The number of
possible serial orders for any execution is not infinite.  In fact,
for each pair of a read and a write with the same variable and a
different value there is exactly one edge in serial order, and this
edge is chosen from two choices.  Therefore, the number of serial
orders for an execution is exactly $2^x$ where $x$ is the number of
such read-write pairs.  When accepting executions, an implementation
of anti-order could be conservative, and only consider a subset of
possible serial orders.  It could even deterministically chose a
single serial order on which to accept executions.  This way, the
implementation could be more efficient without accepting any
unacceptable executions.  However, it might reject some acceptable
executions.  From now on, for purposes of brevity we will use serial
order as if it were a single order.  Any definition using serial order
can be read ``There exists a serial order such that\ldots''

It would be desirable if all four properties were orthogonal, but this
is not the case.  GAO is strictly stronger than GDO which is proven
below.  One goal of this work was to develop GAO to be as weak as
possible while still supporting the assertion that GPO+GDO+GWO+GAO is
equivalent to sequential consistency.  Every candidate definition of
GAO that was not stronger that GDO did not support equivalence to
sequential consistency.  This may reveal some fundamental aspect of
consistency models, or it may merely require further research to
develop such a definition.  As a result, GDO+GAO is equivalent to just
GAO.

\begin{lemma}

\label{not-GDO-not-GAO}

If data order has a cycle, then the execution is not GAO.

\end{lemma}

\begin{quotation}

Proof:

\begin{description}

\item[Case 1: The cycle has a read] Take the operation immediately
before the read in the cycle.  If it is linked by a transitive chain
add that transitive chain to the cycle.  Repeat until the operation
immediately before the read is linked directly without a transitive
chain.  This is either a write, or by lemma~\ref{do-reads} it is a
read ordered by process order.  If it is a read, repeat until a write
is reached.  A write must be reached because otherwise the cycle will
return to the original read which must be ordered before itself by
process order which is a contradiction.  The write that is reached is
directly ordered by data order before the next operation in the cycle
which is a read.  They cannot be ordered by condition~\ref{pre-read}
of data order because this would imply that there exists a third
operation such that the write is process ordered before that
operation, and the read writes to that operation.  This is impossible
since a read cannot write to another operation.  So the write must be
ordered before the read by process order or writes-to order.  Also,
the operations are in a cycle in data order so the read is data
ordered before the write.  In either case, the write is anti ordered
before itself.  The serial views for GAO must all contain this write,
so they cannot respect this cycle in anti order.  Therefore, the
execution is not GAO.

\item[Case 2: The cycle has no reads] Once again, expand the cycle so
that no link is a transitive chain.  If the transitive chain includes
a read refer to case 1.  None of the links can result from writes-to
order because a write cannot write to another write.  The cycle must
contain writes from at least two processes.  If not, a write must be
ordered by data order before another write earlier in process order.
This must have come about by condition~\ref{pre-read} of data order.
Therefore, the following condition exists, $w_1<_{PO}w_2<{PO}r$, and
$w_1\mapsto r$.  all of these operations are to the same variable.  It
is impossible for this process' view to be serial and respect local
order, so the execution is not GAO.  So there must be some links that
result from condition~\ref{pre-read} of data order between writes by
different processes.  Pick one write, $w_1$ and follow the cycle along
process order links until a link resulting form
condition~\ref{pre-read} is reached.  In this case, a write, $w_2$ is
process ordered before a read, $r$, which is written-to by another
write, $w_3$, creating the link $w_2<_{DO}w_3$.  $w_1$ must also be
process ordered before $r$ because either it is process ordered before
$w_2$, or it is $w_2$.  So $w_1<_{DO}w_3$.  Now, $w_1$ does not write
to $r$, so it must be ordered by serial order either
$w_1<_{SO}w_3\mapsto r$, or $r<_{SO}w_1$.  The second case is
impossible.  The view for the process that submitted $w_1$ and $r$
must contain both operations and respect local order.  The assignment
of $r<_{SO}w_1$ would prevent this, and so this assignment could never
be used to show that the execution is GAO.  Therefore, the assignment
must be $w_1<_{SO}w_3$.  By the same logic, follow the chain from
$w_3$ to the next link that results from condition~\ref{pre-read}.
There must be another write serial ordered after $w_3$.  Every time
the cycle switches to an operation by a different process, the first
operation by the new process must be serial ordered after $w_3$.
Continue around the cycle.  At some point the cycle will change
processes for the last time before reaching $w_1$.  The first
operation by this new process is either $w_1$, or a write process
ordered before $w_1$.  This write must also be serial ordered before
$w_3$.  Either it is $w_1$, or it is process ordered before $r$, and
the same reasoning applies.  This assignment of serial order has a
cycle involving only writes, and so no process' view could respect it.
We have previously shown that any alternate assignment would also
prevent the execution from satisfying GAO.  Therefore, the execution
is not GAO.

\end{description}

\end{quotation}

\begin{theorem}

GAO is strictly stronger than GDO.

\end{theorem}

\begin{quotation}

Proof: GAO is shown to be stronger by Theorem~\ref{acyclic} and
Lemma~\ref{not-GDO-not-GAO}.  GAO is shown to be strictly stronger by
the fact that the execution in Figure~\ref{non-sequential} satisfies
GDO and not GAO.

\end{quotation}

All that remains is to show that the four properties together make up
sequential consistency.  Since GAO is stronger than GDO we will leave
it out and prove that GPO+GWO+GAO is equivalent to sequential
consistency.

\begin{lemma}

\label{all-four-if}

Every sequentially consistent execution is GPO+GWO+GAO.

\end{lemma}

\begin{quotation}

Proof: A sequentially consistent execution has a single, serial view
on all operations that respects $<_{PO}$.  Call this view $<_{seq}$.
By definition, $<_{seq}$ respects $<_{PO}$.  If $w_1<_{WO}w_2$ then
$\exists r$ such that $w_1\mapsto r<_{PO}w_2$.  $<_{seq}$ respects
$<_{PO}$ and is serial so it respects $\mapsto$ and therefore it
respects $<_{WO}$.

Now we will show that a sequentially consistent execution respects
$<_{DO}$.  This is not strictly required by the theorem, but will make
it easier to prove that the execution satisfies $<_{AO}$.  $<_{seq}$
respects $<_{PO}$ and is serial, and so respects the $<_{PO}$ and
$\mapsto$ conditions of $<_{DO}$.  If $o_1<_{PO}r$, and $o_2\mapsto
r$, and $o_1$ has a different value than $r$ then $o_1$ must come
before $o_2$ in $<_{seq}$, or the view will not be serial.  If this
were not so then $o_1$ must come between $o_2$ and $r$ because
$o_1<_{PO}r$ and $<_{seq}$ respects $<_{PO}$.  There are two cases,
$o_1$ is either a write or a read.  If $o_1$ is a write then $r$ does
not read from the most recent write and $<_{seq}$ is not serial.  If
$o_1$ is a read then either $o_1$ does not read from the most recent
write, or there is a write to the same variable with the same value as
$o_1$ between $o_2$ and $o_1$ in which case $r$ does not read from the
most recent write and $<_{seq}$ is not serial.  Therefore, $<_{seq}$
respects condition~\ref{pre-read} of $<_{DO}$.  $<_{seq}$ is a total
order.  Since it respects the first three conditions of $<_{DO}$ it
will respect the transitive closure condition.

To prove that a sequentially consistent execution is GAO, define a
serial order, $<_{SO}$, with edges in the same order as $<_{seq}$.
This is possible because if $\exists w,w',r$ such that $w'\mapsto r$
and $w\neq w'$ then it cannot be that $w'<_{seq}w<_{seq}r$ because
then $<_{seq}$ would not be serial.  $w$ must be ordered either before
$w'$ or after $r$.  If $\exists w_1,w_2$ such that $w_1<_{AO}w_2$

then $\exists r_1,r_2$ such that $w_1\mapsto r_1<_{PO}r_2<_{DO}w_2$,
or $w_1\mapsto r_1<_{PO}r_2<_{SO}w_2$, or $w_1\mapsto r_1<_{SO}w_2$,
or $w_1<_{PO}r_1<_{DO}w_2$, or $w_1<_{PO}r_1<_{SO}w_2$.  $<_{seq}$
respects $\mapsto$, $<_{PO}$, $<_{DO}$, and $<_{SO}$ so therefore
respects $<_{AO(<_{SO})}$.

So $<_{seq}$ respects $<_{PO}$, $<_{WO}$, $<_{SO}$, $<_{AO(<_{SO})}$,
is serial, and contains all operations so it can be used to construct
the required per-process views for all processes:

\begin{quote}

$\forall_{i\in
P}\exists$SerialView$(<_{iLocal}\cup<_{PO}\cup<_{WO}\cup<_{SO}\cup<_{AO(<_{SO})}|(*,i,*,*)\cup(w,*,*,*))$

\end{quote}

so the execution is GPO+GWO+GAO.

\end{quotation}

\begin{lemma}

\label{order-writes}

For any GPO+GWO+GAO execution the per-process views can be constructed
where all write operations occur in the same order in all views.

\end{lemma}

\begin{quotation}

Proof: Because $<_{iLocal}$ is a subset of $<_{PO}$ we will ignore it
and just show that the constructed views respect $<_{PO}$, $<_{WO}$,
$<_{SO}$, $<_{AO(<_{SO})}$, and are serial.  There must be an initial
definition of serial order for which the execution satisfies
GPO+GWO+GAO.  This definition of serial order is not changed
throughout this proof.  That is, the final constructed views satisfy
GPO+GWO+GAO for the same definition of serial order as the initial
views.  All initial writes must be ordered first in all views because
all initial writes are ordered before any other operation by $<_{PO}$.
These initial writes can come in any order because they are not
ordered with respect to each other, and there are no reads between
them, so place them in the same order in all views.  For any two views
$<_i$ and $<_j$, the first write that is not an initial write in $<_i$
can be placed first in $<_j$.  Then the next write in $<_i$ can be
placed second in $<_j$, and so on.  We will use an inductive proof to
show that this reordering can be done and the resulting views will
still respect $<_{PO}$, $<_{WO}$, $<_{SO}$, $<_{AO(<_{SO})}$, and be
serial.  The inductive proof uses the following definitions and
invariants:

\begin{enumerate}

\item The order $<$ is defined as
$<_{PO}\cup<_{WO}\cup<_{SO}\cup<_{AO(<_{SO})}$.

\item The views $<_i$ and $<_j$ respect $<$ and are serial.

\item The write operation being moved is called $w_1$.

\item Point A is the place in $<_j$ where $w_1$ will be moved to.

\item Point B is the place in $<_j$ where $w_1$ is being moved from.

\item Point B is after point A in $<_j$.

\item All write operations ordered before $w_1$ in $<_i$ are before
point A in $<_j$.

\item Corollary: All write operations ordered before $w_1$ by $<$ are
before point A in $<_j$ because $<_i$ respects $<$.

\end{enumerate}

The execution is GPO+GWO+GAO so there must exist initial views $<_i$
and $<_j$ that respect $<$ and are serial.  In the initial case, point
A is just after the initial writes of $<_j$.  $w_1$ is the first
non-initial write in $<_i$ so only the initial writes are ordered
before it in $<_i$ and they are all before point A in $<_j$.  $W_1$ is
after the initial writes in $<_j$ so point B is after point A in
$<_j$.

Consider all the operations between A and B.  These must all be either
read operations by process $j$, or write operations not ordered before
$w_1$ by $<$.  Construct the set of prior reads as follows.  The
variable that $w_1$ operates on will be referred to as $x$.  Any read
between A and B to variable $x$ is a prior read.  Also, any read
between A and B ordered by process order before $w_1$ or a prior read
is a prior read.  Then construct the set of remaining operations as
all reads between A and B that are not prior reads plus all writes
between A and B.  Now, we will show that $w_1$ or any prior read can
not be ordered after any remaining operation.

\begin{description}

\item[Case 1: $w_1$ was submitted by process $j$] Every read between A
and B is a prior read.  The remaining operations are all writes and
cannot be ordered by $<$ before $w_1$ by the invariant.  The remaining
operations also cannot be ordered before any prior read by $<$.  They
cannot be ordered by $<_{PO}$ because the write would be by process
$j$ and so would be ordered before $w_1$ which is a contradiction.  A
read and a write cannot be ordered by $<_{WO}$ or $<_{AO(<_{SO})}$
because those two orders only occur between pairs of write operations.
Also, a read cannot be ordered after a write by $<_{SO}$.

\item[Case 2: $w_1$ was not submitted by process $j$] If a remaining
operation is a read it is by process $j$ so it cannot be ordered
before $w_1$ by $<_{PO}$.  The remaining read also cannot be ordered
before $w_1$ by $<_{WO}$ or $<_{AO(<_{SO})}$ because those orders only
occur between pairs of write operations. The remaining read cannot be
ordered before $w_1$ by $<_{SO}$ because the read would be to the same
variable as $w_1$, and so would be a prior read.  The remaining read
cannot be ordered before any prior read because all reads are by
process $j$ so it would be ordered before a prior read by $<_{PO}$
making it a prior read.

If the remaining operation is a write it cannot be ordered by $<$
before $w_1$ by the invariant.  It cannot be ordered before a prior
read by $<_{WO}$ or $<_{AO(<_{SO})}$ because those only order pairs of
writes.  It cannot be ordered before a prior read by $<_{SO}$ because
a read cannot be ordered after a write by $<_{SO}$.  All that remains
is to show that a remaining operation which is a write cannot be
ordered before a prior read by $<_{PO}$.  Any prior read, $r$, comes
before $w_1$ in $<_j$.  The write, $w_2$, which wrote to $r$ must also
come before $w_1$ because $<_j$ is serial.  If $r$ is to the same
variable as $w_1$ then either, $w_1<_{SO}w_2$, or $r<_{SO}w_1$.  Since
$<_j$ respects $<_{SO}$ it must be the case that $r<_{SO}w_1$.  If a
remaining operation $w_3$ is ordered before a prior read, $r_1$, by
$<_{PO}$ then either $r_1$ is to the same variable as $w_1$ in which
case $r_1<_{SO}w_1$, or $r_1$ is ordered by $<_{PO}$ before $r_2$
which is to the same variable as $w_1$ in which case $r_2<_{SO}w_1$.
Therefore, $w_3<_{PO}(r_1<_{PO})r_2<_{SO}w_1$ so $w_3<_{AO}w_1$ which
is a contradiction of the invariant.

\end{description}

In either case, $w_1$ and all prior reads are not ordered after any
remaining operations by $<$.  Now $<_j$ is changed as follows: All
prior reads are placed immediately before point A preserving their
order followed by $w_1$.  All other operations preserve their order.
For all pairs of operations that change their relative position one
must be $w_1$ or a prior read.  The other must be a remaining
operation.  These pairs cannot be ordered by $<$ so the view still
respects $<$.

Before the move, $<_j$ was serial so each prior read must have read
from the most recent write to that variable.  That write must have
been before point A because it is anti ordered before $w_1$.  The
write must still be the most recent write to the same variable because
the moved read is after all writes before point A, and every write
between the two was there before the move when $<_j$ was serial.
Remaining reads maintained their relative position with all writes
except $w_1$.  Remaining reads cannot be to variable $x$, and so they
too must still read from the most recent write.  No other pairs of
reads and writes changed relative position so $<_j$ must still be
serial.

Now, move point A to immediately after $w_1$.  The next write in $<_i$
becomes the new $w_1$.  This write has not been moved to before point
A in $<_j$ so point B is still after point A.  The set of writes
before $w_1$ in $<_i$ have all been moved to before point A in $<_j$,
so the invariants are satisfied.  Therefore, by induction one can
create views for all processes that respect $<$ and have the write
operations in the same order in all views.

\end{quotation}

\begin{lemma}

\label{all-four-only-if}

For any GPO+GWO+GAO execution it is possible to construct a single
view containing all operations that respects process order and is
serial.

\end{lemma}

\begin{quotation}

Proof: From lemma~\ref{order-writes} create views which all have the
write operations in the same order.  These orders respect $<_{PO}$ and
are serial.  Then take one of these views and add the read operations
of all other processes in the same relative position to the writes as
they occur in their own view.  The read operations must all be ordered
by $<_{PO}$ correctly with respect to all writes because the writes
occur in the same order in every view.  Reads ordered with respect to
each other by $<_{PO}$ come from the same view, and so they are placed
in that order in the new view.  The serial property is not affected by
the relative position of pairs of reads, and every read operation is
in the same position relative to all writes, so the view must be
serial.

\end{quotation}

\begin{theorem}

GPO+GWO+GAO is equivalent to sequential consistency.

\end{theorem}

\begin{quotation}

Proof: Follows directly from lemmas~\ref{all-four-if}
and~\ref{all-four-only-if}.

\end{quotation}

Adding GAO almost completes the lattice as shown in
Figure~\ref{fig-lattice}.  Since GAO is stronger than GDO any box labeled
with GAO will also enforce GDO, but that is not shown for brevity.
The lattice now has three additional new consistency models: GAO,
GPO+GAO, and GWO+GAO.  The lattice is almost complete, but it does not
yet contain slow consistency.  Slow consistency would be located below
both PRAM and cache, and above local.

\subsection{Slow Consistency as a Combination of Properties}

In slow consistency~\cite{hutto90:slow}, two operations must maintain
their order only if they are by the same process and to the same
variable.  This leads to the following definitions.

\begin{definition}

Two operations are ordered by \emph{process-data order},
$o_1<_{PDO}o_2$, iff $o_1<_{PO}o_2$, and $o_1<_{DO}o_2$.

\end{definition}

\begin{definition}

An execution is \emph{Global Process-Data Order (GPDO)} iff

\begin{quote}

$\forall_{i\in P}\exists$
SerialView$(<_{iLocal}\cup<_{PDO}|(*,i,*,*)\cup(w,*,*,*))$

\end{quote}

\end{definition}

\begin{theorem}

\label{GPDO}

GPDO is equivalent to slow consistency.

\end{theorem}

\begin{quotation}

Proof: For any GPDO execution, take the view for a single processor.
Divide this view into separate views, one for each variable by
restricting the set of operations to operations on a single variable,
but maintaining their relative order.  Process-data order contains all
edges in process order between operations to the same variable.  These
views respect process-data order, and contain only operations to a
single variable so they respect process order.  These views are
exactly what is required to satisfy slow consistency.

For any slow consistent execution, gather together the views over all
variables for a particular processor.  By similar logic to
Lemma~\ref{GDO-iLocal}, the union of these views and $<_{iLocal}$ must
be acyclic.  The union of the views must contain every edge in
process-data order.  Therefore, any topological sort of the union of
the views and $<_{iLocal}$ must respect $<_{iLocal}\cup<_{PDO}$.
Also, each view is serial.  In the topological sort, every pair of
operations to the same variable must preserve their relative position
so the topological sort must be serial.  The topological sort is
exactly what is required to satisfy GPDO.

\end{quotation}

GPDO is more than just a new statement of slow consistency.  It
represents a new way of combining consistency properties.  We have
already seen GPO+GDO as a way to combine two models to produce a
stronger model.  Now, GPDO combines two models to produce a weaker
model.  This could be done for any pair of properties.  For example,
process-anti order orders only operations that are ordered by both
process order and anti order.  GPAO would be weaker than both GPO and
GAO.  However, it is questionable how useful models this weak would
be.  Slow consistency is essentially only valuable in defining
synchronized models.  Perhaps these models would be usable with a
transition theory, and higher consistency operations between them for
synchronization.

\subsection{A Lattice of Consistency Models}

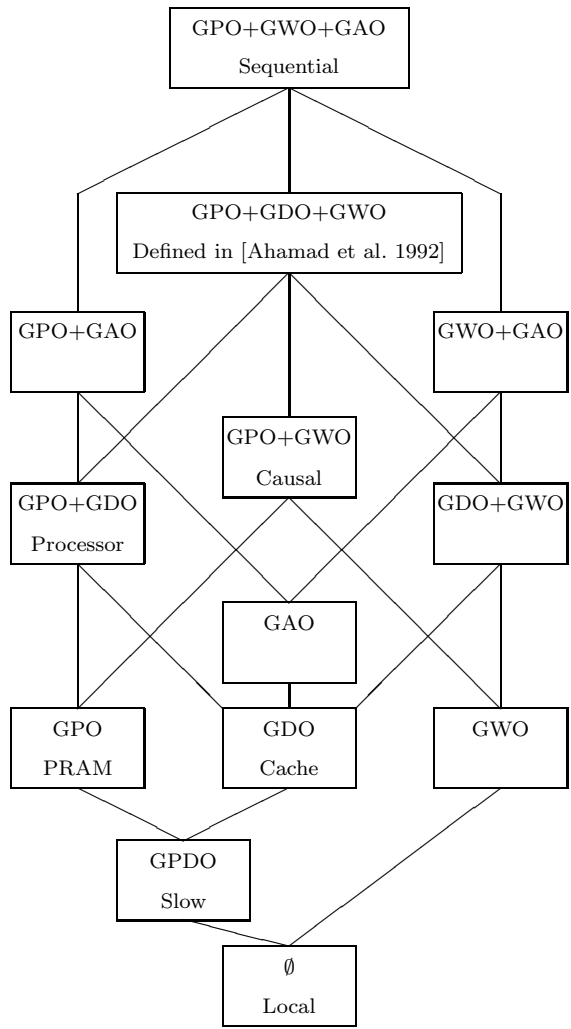
\begin{figure}[tp]

\begin{center}

\begin{picture}(210,385)

\put(60,355){\framebox(90,30){}}
\put(60,370){\makebox(90,15){GPO+GWO+GAO}}
\put(60,355){\makebox(90,15){Sequential}}

\put(105,355){\line(-2,-1){80}}
\put(105,355){\line(0,-1){40}}
\put(105,355){\line(2,-1){80}}

\put(25,315){\line(0,-1){45}}
\put(185,315){\line(0,-1){45}}

\put(0,240){\framebox(50,30){}}
\put(0,255){\makebox(50,15){GPO+GAO}}
\put(40,285){\framebox(130,30){}}
\put(40,300){\makebox(130,15){GPO+GDO+GWO}}
\put(40,285){\makebox(130,15){Defined in~\cite{ahamad92:processor}}}
\put(160,240){\framebox(50,30){}}
\put(160,255){\makebox(50,15){GWO+GAO}}

\put(25,240){\line(0,-1){35}}
\put(25,240){\line(1,-1){80}}
\put(105,285){\line(-1,-1){80}}
\put(105,285){\line(0,-1){55}}
\put(105,285){\line(1,-1){80}}
\put(185,240){\line(-1,-1){80}}
\put(185,240){\line(0,-1){35}}

\put(0,175){\framebox(50,30){}}
\put(0,190){\makebox(50,15){GPO+GDO}}
\put(0,175){\makebox(50,15){Processor}}
\put(80,200){\framebox(50,30){}}
\put(80,215){\makebox(50,15){GPO+GWO}}
\put(80,200){\makebox(50,15){Causal}}
\put(160,175){\framebox(50,30){}}
\put(160,190){\makebox(50,15){GDO+GWO}}

\put(25,175){\line(0,-1){55}}
\put(25,175){\line(1,-1){55}}
\put(105,200){\line(-1,-1){80}}
\put(105,200){\line(1,-1){80}}
\put(185,175){\line(-1,-1){55}}
\put(185,175){\line(0,-1){55}}

\put(80,130){\framebox(50,30){}}
\put(80,145){\makebox(50,15){GAO}}

\put(105,130){\line(0,-1){10}}

\put(0,90){\framebox(50,30){}}
\put(0,105){\makebox(50,15){GPO}}
\put(0,90){\makebox(50,15){PRAM}}
\put(80,90){\framebox(50,30){}}
\put(80,105){\makebox(50,15){GDO}}
\put(80,90){\makebox(50,15){Cache}}
\put(160,90){\framebox(50,30){}}
\put(160,105){\makebox(50,15){GWO}}

\put(25,90){\line(2,-1){40}}
\put(105,90){\line(-2,-1){40}}
\put(185,90){\line(-4,-3){80}}

\put(40,40){\framebox(50,30){}}
\put(40,55){\makebox(50,15){GPDO}}
\put(40,40){\makebox(50,15){Slow}}

\put(65,40){\line(4,-1){40}}

\put(80,0){\framebox(50,30){}}
\put(80,15){\makebox(50,15){$\emptyset$}}
\put(80,0){\makebox(50,15){Local}}

\end{picture}

\end{center}

\caption{The Complete Lattice of Consistency Models}

\label{fig-lattice}

\end{figure}

The result of these composable consistency properties is the lattice
of consistency models shown in Figure~\ref{fig-lattice}.  Every
possible combination of properties produces a model represented by a
box in the lattice.  The top of the lattice is sequential consistency,
and the bottom is local consistency.  Every pair of models has a
unique least upper bound and greatest lower bound.  There are other
combinations of properties demonstrated in this work such as
GPO$\cap$GDO, and GPAO.  These are not shown in the lattice for
brevity, and because their utility is unknown.  GPDO is shown in the
lattice because slow consistency is a well known and widely used
model.

One can think of every box in the lattice as representing a set of
executions that satisfies that model, and no stronger model in the
lattice.  To show that every box of the lattice is non-empty we
provide example executions that violate each of the four consistency
properties.  To derive an example execution for a particular box,
combine the executions violating all the properties not contained in
that box.  Figure~\ref{non-sequential} given when defining anti-order
in Subsection~\ref{anti-property} provides an execution that violates GAO
without violating any of the other three properties.

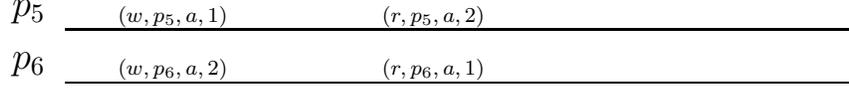
\begin{figure}[tp]

\begin{center}

\begin{picture}(320,40)(0,0)

\process{5}{25}{20}
\put(40,23){$(w,p_5,a,1)$}
\put(140,23){$(r,p_5,a,2)$}
\process{6}{5}{0}
\put(40,3){$(w,p_6,a,2)$}
\put(140,3){$(r,p_6,a,1)$}

\end{picture}

\end{center}

\caption{An Execution That Violates GDO}

\label{non-GDO}

\end{figure}

Figure~\ref{non-GDO} provides an execution that violates GDO (and thus
GAO), but does not violate GPO or GWO.  From condition~\ref{pre-read}
of data order:

\begin{quote}

$(w,p_5,a,1)<_{DO}(w,p_6,a,2)<_{DO}(w,p_5,a,1)$

\end{quote}

Therefore, there is a cycle in $<_{DO}$ so the execution is not GDO.
However, write-read-write order is empty.  The following views satisfy
$<_{PO}$ and $<_{WO}$, and are serial.

\begin{quote}

$p_5:(w,p_5,a,1)<_5(w,p_6,a,2)<_5(r,p_5,a,2)$\\
$p_6:(w,p_6,a,2)<_6(w,p_5,a,1)<_6(r,p_6,a,1)$

\end{quote}

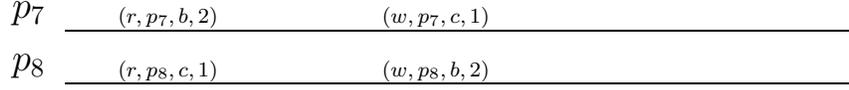
\begin{figure}[tp]

\begin{center}

\begin{picture}(320,40)(0,0)

\process{7}{25}{20}
\put(40,23){$(r,p_7,b,2)$}
\put(140,23){$(w,p_7,c,1)$}
\process{8}{5}{0}
\put(40,3){$(r,p_8,c,1)$}
\put(140,3){$(w,p_8,b,2)$}

\end{picture}

\end{center}

\caption{An Execution That Violates GWO}

\label{non-GWO}

\end{figure}

Figure~\ref{non-GWO} provides an execution that violates GWO, but does
not violate GPO, GDO, or GAO.  The following cycle exists.

\begin{quote}

$(w,p_7,c,1)<_{WO}(w,p_8,b,2)<_{WO}(w,p_7,c,1)$

\end{quote}

These two writes must be present in all views, so no view can respect
$<_{WO}$.  Each write is data ordered before the read it writes-to.
Serial order and anti order are empty.  The following views satisfy
$<_{PO}$, $<_{DO}$, $<_{AO(<_{SO})}$, $<_{SO}$, and are serial.

\begin{quote}

$p_7:(w,p_8,b,2)<_7(r,p_7,b,2)<_7(w,p_7,c,1)$\\
$p_8:(w,p_7,c,1)<_8(r,p_8,c,1)<_8(w,p_8,b,2)$

\end{quote}

To produce an execution that satisfies only GPO and no stronger model
in the lattice, define an execution containing $p_5$ and $p_6$ from
Figure~\ref{non-GDO} and $p_7$ and $p_8$ from Figure~\ref{non-GWO}.
Likewise, to create an execution satisfying only GPO+GDO combine
Figure~\ref{non-sequential} with Figure~\ref{non-GWO}, and so forth.

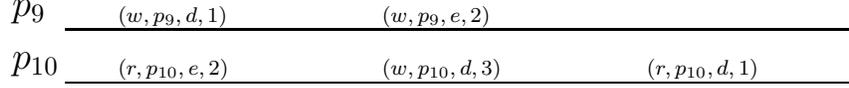
\begin{figure}[tp]

\begin{center}

\begin{picture}(320,40)(0,0)

\process{9}{25}{20}
\put(40,23){$(w,p_9,d,1)$}
\put(140,23){$(w,p_9,e,2)$}
\process{10}{5}{0}
\put(40,3){$(r,p_{10},e,2)$}
\put(140,3){$(w,p_{10},d,3)$}
\put(240,3){$(r,p_{10},d,1)$}

\end{picture}

\end{center}

\caption{An Execution That Violates GPO}

\label{non-GPO}

\end{figure}

Figure~\ref{non-GPO} provides an execution that violates GPO, but does
not violate GDO, GWO, or GAO.  In order for the view for $p_{10}$ to
be serial, $(w,p_9,e,2)$ must come before $(r,p_{10},e,2)$, and
$(w,p_9,d,1)$ must come after $(w,p_{10},d,3)$.  In order to respect
local order, $(r,p_{10},e,2)$ must come before $(w,p_{10},d,3)$.
Therefore, $(w,p_9,e,2)$ must come before $(w,p_9,d,1)$ which does not
respect $<_{PO}$.

The following are the definitions of $<_{DO}$ and $<_{WO}$ for this
execution.

\begin{quote}

$(w,p_{10},d,3)<_{DO}(w,p_9,d,1)<_{DO}(r,p_{10},d,1)$\\
$(w,p_9,e,2)<_{DO}(r,p_10,e,2)$\\
$(w,p_9,e,2)<_{WO}(w,p_{10},d,3)$

\end{quote}

With the following definition of serial order, anti order is empty.

\begin{quote}

$(w,p_{10},d,3)<_{SO}(w,p_9,d,1)$

\end{quote}

The following view for $p_{10}$ satisfies $<_{DO}$, $<_{WO}$,
$<_{AO(<_{SO})}$, $<_{SO}$, $<_{p_{10}Local}$, and is serial.

\begin{quote}

$p_{10}:(w,p_9,e,2)<_{10}(r,p_{10},e,2)<_{10}(w,p_{10},d,3)<_{10}(w,p_9,d,1)<_{10}(r,p_{10},d,1)$

\end{quote}

However, the view for $p_9$ is not as simple.  The following cycle
exists.

\begin{quote}

$(w,p_{10},d,3)<_{SO}(w,p_9,d,1)<_{p_9Local}(w,p_9,e,2)<_{WO}(w,p_{10},d,3)$

\end{quote}

No view can be written for $p_9$ that satisfies GWO+GAO.  However,
separate views can be written, one that satisfies GWO, and one that
satisfies GAO.

\begin{quote}

$p_9(GWO):(w,p_9,d,1)<_9(w,p_9,e,2)<_9(w,p_{10},d,3)$\\
$p_9(GAO):(w,p_{10},d,3)<_9(w,p_9,d,1)<_9(w,p_9,e,2)$

\end{quote}

Therefore, this execution satisfies GAO, and no stronger model in the
lattice.  It also satisfies GWO, and no stronger model in the lattice.
By combining this execution with Figure~\ref{non-sequential} we
achieve an execution that satisfies only GDO.  All that remains is to
find executions that satisfy GWO+GAO and GDO+GWO.

\begin{figure}[tp]

\begin{center}

\begin{picture}(320,40)(0,0)

\process{11}{25}{20}
\put(40,23){$(w,p_{11},f,1)$}
\put(140,23){$(w,p_{11},g,4)$}
\put(240,23){$(r,p_{11},g,3)$}
\process{12}{5}{0}
\put(40,3){$(w,p_{12},g,3)$}
\put(140,3){$(w,p_{12},f,2)$}
\put(240,3){$(r,p_{12},f,1)$}

\end{picture}

\end{center}

\caption{An Execution That Satisfies GWO+GAO}

\label{GWO+GAO}

\end{figure}
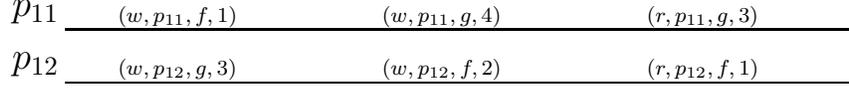

Figure~\ref{GWO+GAO} satisfies GWO+GAO, but not GPO+GWO+GAO.  Below is
the definition of $<_{DO}$.

\begin{quote}

$(w,p_{12},f,2)<_{DO}(w,p_{11},f,1)<_{DO}(r,p_{12},f,1)$\\
$(w,p_{11},g,4)<_{DO}(w,p_{12},g,3)<_{DO}(r,p_{11},g,3)$

\end{quote}

The following definition of serial order must be chosen.

\begin{quote}

$(w,p_{11},g,4)<_{SO}(w,p_{12},g,3)$

\end{quote}

If not then $(w,p_{11},g,4)$ must be ordered after $(r,p_{11},g,3)$
which violates the order $<_{p_{11}Local}$.  Likewise for
$(w,p_{12},f,2)$ and $(r,p_{12},f,1)$.  $<_{WO}$ and $<_{AO(<_{SO})}$
are empty.  The following cycle exists.

\begin{quote}

$(w,p_{11},g,4)<_{SO}(w,p_{12},g,3)<_{PO}(w,p_{12},f,2)<_{SO}(w,p_{11},f,1)<_{PO}(w,p_{11},g,4)$

\end{quote}

Therefore, it is impossible for any view to respect both $<_{PO}$ and
$<_{SO}$.  So the execution is not GPO+GAO, and hence it is not
GPO+GWO+GAO.  However, this execution is GWO+GAO as the following
views demonstrate.

\begin{quote}

$p_{11}:(w,p_{12},f,2)<_{11}(w,p_{11},f,1)<_{11}(w,p_{11},g,4)<_{11}$\\
$(w,p_{12},g,3)<_{11}(r,p_{11},g,3)$\\
$p_{12}:(w,p_{11},g,4)<_{12}(w,p_{12},g,3)<_{12}(w,p_{12},f,2)<_{12}$\\
$(w,p_{11},f,1)<_{12}(r,p_{12},f,1)$

\end{quote}

To create an execution that satisfies GDO+GWO and no stronger model
combine this execution with Figure~\ref{non-sequential}.  The complete
lattice as shown in Figure~\ref{fig-lattice} is a powerful new way to
describe and organize consistency models.  Every non-synchronized
model described in Section~\ref{related} is encompassed by the lattice
model.  In addition, five new consistency models are uncovered by the
symmetry of the lattice.  Every model in the lattice has a non-empty
set of executions which satisfy that model and no stronger model in
the lattice.  Finally, new consistency properties would be easy to
integrate into the lattice if they are discovered.  Synchronized
models are not covered directly by the lattice.  Instead, synchronized
models can be viewed as processes submitting some operations under one
consistency model, and some operations under another consistency
model, i.e. a consistency transition.  Synchronized models will
covered in Section~\ref{transitions} on consistency transitions.  The
lattice model facilitates the definition of consistency transitions
because any two models are easily compared by the properties they
enforce.

\section{Consistency Transitions}

\label{transitions}

Our final generalization of consistency models is the idea of
consistency transitions.  In synchronized consistency models, a
program executes ordinary operations with a relaxed consistency model,
usually slow consistency.  Occasionally, the program executes
synchronization operations with a stronger consistency model, usually
sequential consistency.  These synchronization operations enforce
additional ordering restrictions between ordinary operations.  This
can be viewed as a consistency transition where the process executing
a synchronization operation temporarily requests a stronger level of
consistency.  Our goal is to develop a general theory of consistency
transitions between any two consistency models, not just slow and
sequential.  Synchronized models require the following.

\begin{enumerate}

\item All synchronization operations must be sequentially consistent.

\item All ordinary operations must be slow consistent.

\item The order $<_D$ must be respected between synchronization and
ordinary operations

\end{enumerate}

Sequential consistency is equivalent to GPO+GWO+GAO.  So the first
condition can be satisfied with serial views on synchronization
operations.

\begin{quote}

$\forall_{i\in P}\exists$
SerialView$(<_{iLocal}\cup<_{PO}\cup<_{WO}\cup<_{SO}\cup<_{AO(<_{SO})}|$\\
$(sr,i,*,*)\cup(sw,*,*,*))$

\end{quote}

Weak consistency does not include acquire and release operations.
Instead, synchronization operations are special read and write
operations.  To distinguish them we use the operation types $sr$ for
synchronized read and $sw$ for synchronized write.  Remember that for
other synchronized models the writes-to relation is defined with
acquire operations treated as reads, and release operations treated as
writes.  If an acquire is defined as an $sr$ and a release as an $sw$
this definition is equally valid for every synchronized model.

Slow consistency is equivalent to GPDO.  So the second condition can
be satisfied by serial views on ordinary operations.

\begin{quote}

$\forall_{i\in P}\exists$
SerialView$(<_{iLocal}\cup<_{PDO}|(or,i,*,*)\cup(ow,*,*,*))$

\end{quote}

The operation type $or$ is used for ordinary read, and $ow$ for
ordinary write.  The views for synchronization and ordinary operations
are very similar.  They each have one view per processor, and each
view contains the reads of that processor plus all writes.  It would
be nice to combine these views into a single view for each processor
containing both synchronization and ordinary operations.  The view
would have to respect the ordering among synchronization operations,
$<_{synch}$,

\begin{quote} 

$<_{synch}\equiv<_{PO}\cup<_{WO}\cup<_{SO}\cup<_{AO(<_{SO})}|(sr,i,*,*)\cup(sw,*,*,*)$

\end{quote}

and the ordering among ordinary operations, $<_{ord}$,

\begin{quote}

$<_{ord}\equiv<_{PDO}|(r,i,*,*)\cup(w,*,*,*)$

\end{quote}

and it would have to respect $<_{iLocal}$ and $<_D$.  However, this
straightforward approach has some problems.

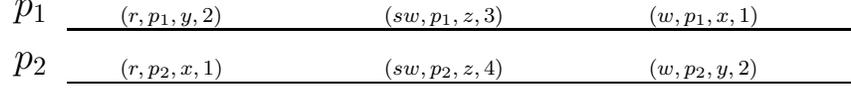
\begin{figure}[tp]

\begin{center}

\begin{picture}(320,40)(0,0)

\process{1}{25}{20}
\put(40,23){$(r,p_1,y,2)$}
\put(140,23){$(sw,p_1,z,3)$}
\put(240,23){$(w,p_1,x,1)$}
\process{2}{5}{0}
\put(40,3){$(r,p_2,x,1)$}
\put(140,3){$(sw,p_2,z,4)$}
\put(240,3){$(w,p_2,y,2)$}

\end{picture}

\end{center}

\caption{An Execution that violates weak consistency}

\label{not-weak}

\end{figure}

Figure~\ref{not-weak} satisfies all of these properties and still does
not satisfy weak consistency as the following views demonstrate.

\begin{quote}

$p_1:(sw,p_2,z,4)<_1(w,p_2,y,2)<_1(r,p_1,y,2)<_1(sw,p_1,z,3)<_1(w,p_1,x,1)$\\
$p_2:(sw,p_1,z,3)<_2(w,p_1,x,1)<_2(r,p_2,x,1)<_2(sw,p_2,z,4)<_2(w,p_2,y,2)$

\end{quote}

Notice that the synchronized writes are unordered by $<_{synch}$.
They may occur in either order, but in this execution they are seen to
occur in different orders by different processes.  Does this violate
the assertion that synchronization operations must be sequentially
consistent?  After all, the synchronization operations by themselves,
ignoring ordinary operations, are sequentially consistent.  The reason
for this conundrum comes from a slight discrepancy between the
intuitive definition and the formal definition of sequential
consistency.  The intuitive definition can be stated like this.

\begin{quote}

There is a single total order of events, and all processes agree that
the events happened in that order.

\end{quote}

However, the formal definition requires that there exist at least one
order of events that every process can agree on.  There may be more
than one order of events that would satisfy every process, and there
is no way to distinguish a single correct order from the sequentially
consistent operations alone.  This problem is not an artifact of our
definition of GPO+GWO+GAO.  It can still occur with the original
definition of sequential consistency.  Below is a restatement of the
definition given previously for synchronized consistency models except
that the positions of $\forall_{i\in P, x\in V}$ and
$\exists<_{seq}=\ldots$ have been reversed.

\begin{quote}

$\forall_{i\in P, x\in
V}\exists<_{seq}=$SerialView$(<_{PO}|(s,*,*,*))$, and\\ $<_S=$the
transitive closure of $<_D\cup<_{seq}$, and\\ $\exists$
SerialView$(<_S\cup<_{PO}|(*,i,x,*)\cup(w,*,x,*))$

\end{quote}

The syncronized operations are sequentially consistent, but each
process gets to choose it's own definition of $<_{seq}$.  This causes
the same problem.  The original definition resolved this problem by
requiring that the definition of $<_S$ for every process be based on a
single definition of $<_{seq}$.  This same strategy can be used with
GPO+GWO+GAO to generate the definition given below.  Note that all
synchronized reads must be included in every view.  This will be
addressed later.

\begin{theorem}

The following definition is equivalent to synchronized model
consistency

\begin{quote}

$\forall_{i\in P}\exists$
SerialView$(<_{iLocal}\cup<_{synch}\cup<_{ord}\cup<_D|(or,i,*,*)\cup$\\
$(ow,*,*,*)\cup(sr,*,*,*)\cup(sw,*,*,*))$ and all synchronization
operations appear in the same order in every view.

\end{quote}

\end{theorem}

\begin{quotation}

Proof: For an execution that satisfies the above views, construct the
original definition synchronized consistent views as follows.  The
view $<_{seq}$ is the total order on synchronization operations that
occurs in every view.  Any two ordinary operations ordered by $<_S$
must be ordered by a transitive chain containing only synchronization
operations.  The per-process views contain all synchronization
operations and respect $<_D$ and $<_{seq}$ so they must also respect
$<_S$.  Construct the per-process per-variable slow consistent views
required by weak consistency from the per-process GPDO consistent
views as shown in Theorem~\ref{GPDO}.  The new views will respect the
old views so they will respect $<_S$.

For an execution that satisfies the original definition of
synchronized consistency, construct the above views as follows.  Begin
with all the synchronization operations in the order specified by
$<_{seq}$.  This is the order in which they will appear in every
per-process view.  The synchronization operations must respect
$<_{synch}$ because by Lemma~\ref{all-four-if} every sequentially
consistent view respects process order, write-read-write order, serial
order, and anti order.  Any single per-process, per-variable slow
consistent view can always be combined with with the synchronization
operations in a way that respects $<_D$ because the view respects
$<_S$ which is the transitive closure of $<_{seq}$ and $<_D$.  Combine
all slow consistent views with the synchronization operations in this
way ignoring, for now, the order between operations from different
slow consistent views.  The resulting view will respect $<_{synch}$,
$<_{ord}$, and $<_D$.  All that remains is to show that it respects
$<_{iLocal}$.

Between two synchronization operations, the ordinary operations can
always be rearranged as a topological sort of $<_{ord}\cup<_{iLocal}$
which is acyclic by Lemma~\ref{GDO-iLocal}.  Two ordinary operations
separated by synchronization operations cannot be out of order with
respect to $<_{iLocal}$ because

\begin{quote}

$o_1<_Ds_1<_{seq}s_2<_Do_2<_{iLocal}o_1$

\end{quote}

This implies that $s_2$ is process ordered before $s_1$, but appears
after it in $<_{seq}$ which is a contradiction.

\end{quotation}

Should all processes be required to see the same total order of
synchronization operations, or is it sufficient that the
synchronization operations are sequentially consistent?  We argue that
sequentially consistency of synchronization operations should be
sufficient even if this allows different processes see different total
orders.  First, we feel that the intuitive definition is in fact
enforcing a consistency model stronger than sequential.  For example,
linearizability~\cite{herlihy90:linear} assumes the existence of a
global Newtonian clock.  The processes may not have access to this
clock, but it does exist.  Each operation spans a certain period of
time.  A linearizable execution must be sequential, and in addition if
two operations have non-overlapping time spans they must appear in the
sequential view in that order.  Perhaps this problem would be solved
if synchronization operations were linearizable, and ordinary
operations had defined time spans and were forced to respect certain
linearizable restrictions with synchronization operations.

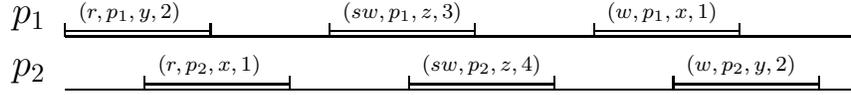
\begin{figure}[tp]

\begin{center}

\begin{picture}(320,40)(0,0)

\process{1}{25}{20}
\put(25,27){$(r,p_1,y,2)$}
\put(20,22){\line(1,0){55}}
\put(20,20){\line(0,1){5}}
\put(75,20){\line(0,1){5}}
\put(125,27){$(sw,p_1,z,3)$}
\put(120,22){\line(1,0){55}}
\put(120,20){\line(0,1){5}}
\put(175,20){\line(0,1){5}}
\put(225,27){$(w,p_1,x,1)$}
\put(220,22){\line(1,0){55}}
\put(220,20){\line(0,1){5}}
\put(275,20){\line(0,1){5}}
\process{2}{5}{0}
\put(55,7){$(r,p_2,x,1)$}
\put(50,2){\line(1,0){55}}
\put(50,0){\line(0,1){5}}
\put(105,0){\line(0,1){5}}
\put(155,7){$(sw,p_2,z,4)$}
\put(150,2){\line(1,0){55}}
\put(150,0){\line(0,1){5}}
\put(205,0){\line(0,1){5}}
\put(255,7){$(w,p_2,y,2)$}
\put(250,2){\line(1,0){55}}
\put(250,0){\line(0,1){5}}
\put(305,0){\line(0,1){5}}

\end{picture}

\end{center}

\caption{Linearizability for Synchronization Operations}

\label{linear}

\end{figure}

For example, Figure~\ref{linear} shows how linearizability could solve
this problem for Figure~\ref{not-weak}.  Even if the time spans for
$(sw,p_1,z,3)$ and $(sw,p_2,z,4)$ overlap, i.e. they can be seen in
either order, there is no way that $(r,p_1,y,2)$ and $(w,p_2,y,2)$ can
overlap while $(r,p_2,x,1)$ and $(w,p_1,x,1)$ also overlap.  The
definitions given for synchronized consistency models explicitly state
that synchronization operations must be sequentially consistent.
However, the implementations given with those definitions implicitly
enforce linearizability over synchronization operations.  The authors
of the various models did not appreciate the effect of this slight
distinction.

Another reason not to require every process to see the same total
order is once again the argument over the distinction between memory
model and programming model.  The reader may have noticed that
Figure~\ref{not-weak} does not implement any kind of mutual exclusion
or barrier behavior.  The program does not know in which order the
synchronized writes occurred, but is relying on the fact that they
occurred in the same order at all processes.  If the program knows
that two synchronization operations occurred in a particular order the
problem disappears.  If the operations are ordered by $<_{synch}$ then
they must appear in that order in all views.  In our opinion, if the
programmer needs two operations to occur in the same order in all
views then the control and data flow of the program must be able to
detect in what order they occurred.  This is part of the programming
model, not the consistency model.  In particular, this problem does
not occur for data-race-free programs because every pair of
conflicting ordinary operations is separated by synchronization
operations with control or data dependencies.  I.e. the
synchronization operations must be ordered by $<_{synch}$.  We propose
to re-define $<_S$ for synchronized consistency models.  Rather than
being the transitive closure of $<_D\cup<_{seq}$ it should be the
transitive closure of $<_D\cup<_{synch}$.  Essentially, the
synchronization operations must be sequentially consistent, and if the
program can tell that two synchronization operations happened in a
particular order then they must be placed in that order in all
process' views.  This leads to a revised definition of synchronized
model consistency.

\begin{definition}

For a given definition of $<_D$, an execution is \emph{synchronized
model consistent} with the new definition $<_S$ iff

\begin{quote}

$\exists<_{seq}=$SerialView$(<_{PO}|(s,*,*,*))$, and\\$<_S=$the
transitive closure of $<_D\cup<_{synch}$, and\\ $\forall_{i\in P, x\in
V}\exists$ SerialView$(<_S\cup<_{PO}|(*,i,x,*)\cup(w,*,x,*))$

\end{quote}

\end{definition}

\begin{theorem}

\label{equivalent-synch}

The following definition is equivalent to synchronized model
consistency with the new definition $<_S$

\begin{quote}

$\forall_{i\in P}\exists$
SerialView$(<_{iLocal}\cup<_{synch}\cup<_{ord}\cup<_D|(or,i,*,*)\cup$\\
$(ow,*,*,*)\cup(sr,*,*,*)\cup(sw,*,*,*))$

\end{quote}

\end{theorem}

\begin{quotation}

Proof: For an execution that satisfies the above views, construct the
synchronized consistent views as follows.  The order $<_{seq}$ is
taken from any of the views as they all contain all synchronization
operations.  Any two ordinary operations ordered by $<_S$ must be
ordered by a transitive chain containing only synchronization
operations.  The per-process views contain all synchronization
operations and respect $<_D$ and $<_{synch}$ so they must also respect
$<_S$.  Construct the per-process per-variable slow consistent views
required by weak consistency from the per-process GPDO consistent
views as shown in Theorem~\ref{GPDO}.  The new views will respect the
old views so they will respect $<_S$.

For an execution that satisfies the new definition of synchronized
consistency, construct the above views as follows.  There must be at
least one order of synchronization operations that respects
$<_{synch}$ because $<_{seq}$ exists.  Furthermore, each per-process,
per-variable slow consistent view respects $<_S$ so it can always be
combined with an ordering of synchronization operations that respects
$<_{synch}$ and $<_D$.  By similar logic as above, combine all
operations for a single process into a single view, and the view will
respect $<_{iLocal}$.

\end{quotation}

Now we will deal with the fact that every synchronized read must be
placed in every view.  The proof above relies on the fact that if
$o_1<_So_2$ then $o_1$ and $o_2$ must be placed in that order in every
view in which they both occur.  This is enforced by the fact that
every view contains all synchronization operations and respects $<_D$
and $<_{synch}$.  If some view were not to contain some synchronized
reads this might not hold.  There are two cases in which ordinary
operations can be ordered by $<_S$.  Case 1, $o_1$ and $o_2$ are
linked by a transitive chain containing at least one $sw$.  In this
case, the $sw$ will be in every view so we can just link the ordinary
operations to the synchronized write instead of any possible
synchronized reads in the chain.  Case 2, $o_1$ and $o_2$ are linked
by a transitive chain containing only synchronized reads.  In this
case, we can link the ordinary operations to each other.  This will be
called transitive order, $<_T$.  In Definition~\ref{transitive},
$<_{synch}^+$ refers to traversing one or more edges of $<_{synch}$.

\begin{definition}

\label{transitive}

Transitive order, $<_T$, is defined as

\begin{quote}

if $o<_Dsr<_{synch}^+sw$ then $o<_Tsw$ \\
if $sw<_{synch}^+sr<_Do$ then $sw<_To$ \\
if $o_1<_Dsr<_Do_2$ then $o_1<_To_2$   \\
if $o_1<_Dsr_1<_{synch}^+sr_2<_Do_2$ then $o_1<_To_2$

\end{quote}

\end{definition}

Now we have another equivalent definition of synchronized model
consistency where each per-process view contains only it's own reads
whether ordinary or synchronized.

\begin{theorem}

The following definition is equivalent to synchronized model
consistency with the new definition $<_S$

\begin{quote}

$\forall_{i\in P}\exists$
SerialView$(<_{iLocal}\cup<_{synch}\cup<_{ord}\cup<_D\cup<_T|(or,i,*,*)\cup(ow,*,*,*)\cup(sr,i,*,*)\cup(sw,*,*,*))$

\end{quote}

\end{theorem}

\begin{quotation}

Proof: By Lemma~\ref{all-four-only-if} it must still be possible to
construct the view $<_{seq}$.  Also, the views must still respect
$<_S$ because any transitive chain in $<_D$ and $<_{synch}$ must be
reflected in the operations present in each view through $<_D$,
$<_{synch}$, and $<_T$.

\end{quotation}

Now this definition can be generalized.  The definition says that
sequential consistency operations must be sequentially consistent with
each other, slow consistent operations must be slow consistent with
each other, and operations of different consistency levels must
respect $<_D$ and $<_T$ between them.  There is no reason this
definition has to be limited to sequential and slow consistency, or
limited to just two consistency levels.  Each operation can be
submitted under a different consistency model; any model within the
lattice.  This leads to a generalized definition of memory
consistency.  Each operation is considered to be labeled with a subset
of the consistency properties, and two operations must respect an
order such as process order if they are both labeled with the global
process order property.

\begin{definition}

Two operations are ordered by \emph{synchronization order}
$o_1<_{synch}o_2$ iff

\begin{quote}

both are labeled GPO and $o_1<_{PO}o_2$, or \\
both are labeled GDO and $o_1<_{DO}o_2$, or \\
both are labeled GWO and $o_1<_{WO}o_2$, or \\
both are labeled GAO and $o_1<_{SO}o_2$ or $o_1<_{AO(<_{SO})}o_2$, or\\
both are labeled GPDO and $o_1<_{PDO}o_2$\ldots

\end{quote}

\end{definition}

\begin{definition}

For a given definition of $<_D$, an execution satisfies
\emph{generalized memory consistency} iff

\begin{quote}

$\forall_{i\in P}\exists$
SerialView$(<_{iLocal}\cup<_{synch}\cup<_D\cup<_T|(r,i,*,*)\cup(w,*,*,*))$

\end{quote}

\end{definition}

So a consistency model is defined by specifying $<_D$ and labeling
operations with consistency properties.  To simulate the
non-synchronized models, $<_D$ is empty and all operations are labeled
with the consistency properties of that model.  $<_{synch}$ reduces to
the union of the orders representing the labeled properties.  For
example, if all operations are labeled GPO+GWO, this definition
reduces to the original definition of causal consistency.  To simulate
the synchronized consistency models, use $<_D$ given for that model.
Synchronization operations are labeled with GPO+GWO+GAO, and ordinary
operations are labeled with GPDO.  This definition can also
accommodate the variant of release consistency where synchronization
operations respect processor consistency.  To simulate location
consistency all that is needed is to replace SerialView with
SerialPartialView as given in Definition~\ref{serialpartialview}.

\begin{figure}[tp]

\begin{center}

\begin{tabular}{ll}

Process $p_1$& Process $p_2$\\
\hline
\multicolumn{2}{l}{The initial value of x is 0;}\\
\\
y = f(input);\\
x = 1; \emph{synch}\\
&while(x==0) wait; \emph{synch}\\
&read(y);

\end{tabular}

\end{center}

\caption{A Data-Race-Free Program}

\label{data-race-free}

\end{figure}

These new definitions also allow new ideas about what it means for a
program to be data-race-free~\cite{adve93:data-race-free}.  A
data-race-free program is one that will only produce sequential
executions even when the memory system supports a particular
consistency model weaker than sequential consistency.  For example,
Figure~\ref{data-race-free} contains a program that will only produce
sequential executions when it is run under weak consistency.  This
program is said to be weak-sequential data-race-free.  A program may
be data-race-free for some non-sequential consistency models, and not
data-race-free for others~\cite{gharachorloo90:release}.  The
operations on x are synchronization operations.  In order to exit the
loop, $p_2$ must read 1 from x.  Therefore, the following ordering
restrictions exist.

\begin{quote}

$(w,p_1,y,f(input))<_D(w,p_1,x,1)\mapsto(r,p_2,x,1)<_D(r,p_2,y,?)$

\end{quote}

The view for $p_2$ must contain all of these operations.  If weak
consistency is enforced, then $<_D$ must be respected, and $\mapsto$
must be respected because the view is serial.  There are no other
writes to y, so $(r,p_2,y,?)$ must return the value written by
$(w,p_1,y,f(input))$.  If this value is returned then the execution is
also sequentially consistent.  One goal of synchronized consistency
models is to simulate sequential consistency in this manner.  This
work provides a new, formal definition of what it means to be a
data-race-free program.  A program is \emph{data-race-free} if and
only if, for any execution produced by the program,

\begin{quote}

Given the definition of $<_D$ and labeling of operations required for
weak consistency:\\$\exists<_{SO}\forall
i\exists$SerialView$(<_{iLocal}\cup<_{synch}\cup<_D\cup<_T|(*,i,*,*)\cup$\\
$(w,*,*,*))$

implies

$\exists<_{SO}\forall
i\exists$SerialView$(<_{PO}\cup<_{WO}\cup<_{AO(<_{SO})}\cup<_{SO}|(*,i,*,*)\cup(w,*,*,*))$

\end{quote}

This literally says that if the program produces a weak consistent
execution, then that same execution is also sequentially consistent.
If the program is run in an environment that only produces weak
consistent executions, then the program will only produce sequentially
consistent executions.  This definition of data-race-free is very
general, but may not be too helpful to programmers.  It does not give
insight on how to write a program that satisfies the condition, and it
may be hard to prove that a particular program satisfies the
condition.  For example, it does not even require that the same
definition of serial order be used to produce the weak consistent
views as the sequentially consistent views.  One could provide
simpler, conservative definitions that are easier to implement and
prove, but still enforce the above condition.  For example, if every
pair of operations ordered by
$<_{PO}\cup<_{WO}\cup<_{AO(<_{SO})}\cup<_{SO}$ were also ordered by
$<_{iLocal}\cup<_{synch}\cup<_D\cup<_T$ then the condition would hold.
A further restriction along these lines is to say that every pair of
ordinary operations to the same variable must be separated by control
and data dependencies among synchronization operations which is the
traditional definition of data-race-free.  This new uniform notation
may allow more precise, less conservative formulations of the class of
data-race-free programs.

\section{Conclusions and Future Work}

\label{conclusion}

The thesis of this work is that every useful shared memory consistency
model (well known and often used models in the literature) can be
described by a single unifying framework.  This work presents such a
framework in the form of a lattice of primitive consistency
properties, and a theory of transitions within the lattice.  Shared
memory can be viewed as an abstract API of interprocess communication
parameterized by its consistency model.  This API can be implemented
in environments with physically shared memory banks in hardware.  Or
in environments with no physically shared memory, as in distributed
shared memory systems.  This style of interprocess communication is
appropriate for many types of applications which can leverage research
done on memory implementations and memory consistency models.

The first contribution of this work is the discovery of four
fundamental consistency properties.  Global Process Order enforces the
condition that all operations by a single process are seen everywhere
in the system to occur in the order in which they were submitted.
Global Data Order enforces the condition that for each variable, there
exists at least one total order of operations which every process can
agree could have been the actual order of those operations.  Combining
these two orders produces another consistency model, GPO+GDO, very
similar to processor consistency.  The difference arises in the fact
that there may be more than one possible total order on each variable
which satisfies data order.  However, data order can be augmented to
be a total order on operations to each variable.  Processor
consistency is equivalent to process order plus this augmented data
order.  This method of combining consistency properties is a general
method which can be used to create a lattice of consistency models.
Any two properties can be combined in this way to produce a
consistency model stronger than either property alone.  This work has
also identified another combination operator which produces a new
model weaker than either property alone.  In this case, GPDO produces
slow consistency.  Thus, all possible combinations of consistency
properties produce a lattice of models.

The third property, Global Write-read-write Order enforces aspects of
causality.  It is defined such that GPO+GWO is equivalent to causal
consistency.  It is the weakest property (the smallest set of edges)
for which this is true.  The fourth property is Global Anti Order.
Anti order is defined such that all four properties combined produce a
model equivalent to sequential consistency.  To accomplish this,
Global Anti Order requires two ordering relations among operations,
anti order and serial order.  Serial order captures the restriction
that every read must read from the most recent write.  Anti order is
based on both serial order and data order.  This complexity is
required as any weaker definition of anti order was not sufficient to
enforce equivalence to sequential consistency.  Another side effect of
this complexity is that Global Anti Order is not orthogonal to all
three other properties.  It is strictly stronger than data order.

The second contribution of this work is the concept of a consistency
lattice.  As stated before, enumerating every combination of the four
consistency properties with both combination operators produces a
lattice of consistency models.  The strongest model in the lattice is
sequential consistency, and the weakest is local consistency.  Every
non-synchronized consistency model described in
Subsection~\ref{non-synch} is equivalent to a node in this lattice.  The
lattice model validates the derived consistency properties as
necessary and sufficient to describe all such models.  Furthermore,
for every consistency model in the lattice there exists a non-empty
set of executions accepted by that model and no stronger model in the
lattice.

The third contribution of this work is that the lattice includes five
previously unnamed, non-empty consistency models: GWO, GAO, GDO+GWO,
GPO+GAO, GWO+GAO.  We believe the most promising of these is GDO+GWO.
It is a data-centric version of causality where operations are placed
in causal order when they are applied to their variable, not when
they are issued by their process.

The fourth contribution of this work is a transition theory over the
consistency lattice.  The uniform lattice framework assists in the
development of the transition theory because any two models can be
compared by their properties, and transitions can be viewed as adding
or removing properties.  The transition theory was evaluated against
synchronized consistency models, and every synchronized model
described in Sec~\ref{synch} can be modeled by this transition theory.
This led to the development of a single statement of consistency
called generalized consistency.  Under generalized consistency, every
operation is labeled with a set of consistency properties.
Consistency requirements among operations depend on their labelings.
If every operation is labeled with the same set of properties,
generalized consistency simulates the non-synchronized consistency
model represented by the combination of those properties.  Various
other labelings simulate the transitions equivalent to the
synchronized models.

In the future, this work can be extended in several directions.  In
the lattice, the five new consistency models need to be examined to
determine intuitive definitions of the effects enforced by those
models, and whether existing applications may be able to take better
advantage of the new models.  The space of consistency models around
processor consistency needs to be explored in more detail as well as
other methods of combining properties such as GPO$\cap$GDO and GPAO.
Finally, efficient implementations could be examined with regards to
what consistency properties they enforce.  A lattice of
implementations related to the lattice of consistency models would be
helpful in automating selection of memory implementations.

\begin{received}
Received Month Year; revised Month Year; accepted Month Year
\end{received}

\bibliographystyle{acmtrans}
\bibliography{dsm}

\appendix
\section*{APPENDIX}
\setcounter{section}{1}

\begin{definition}

An \emph{execution} is a set of \emph{processes}, $P$, a set of
\emph{shared variables}, $V$, a set of \emph{operations}, $O$, and two
partial orders on $O$, \emph{process order}, $<_{PO}$, and
\emph{writes-to order}, $\mapsto$.

\end{definition}

\begin{definition}

An \emph{operation} is a tuple $(op,i,x,v)$ where $op$ is $r$ for a
read, $w$ for a write, or $o$ if the type of operation is unknown.
$i\in P$ is the process submitting the operation.  $x\in V$ is the
variable to which the operation is applied, and $v$ is a valid value
for the variable $x$.

\end{definition}

\begin{definition}

An \emph{operation pattern} is written like an operation with $*$ in
place of one or more of the attributes.  It represents the set of all
operations in $O$ that match the pattern in all attributes that are
not $*$.

\end{definition}

For example, $(r,p_1,x,5)$ denotes that process $p_1$ read the
variable $x$, and received the value $5$.  $(w,*,*,*)$ denotes the set
of all write operations.

\begin{definition}

The set of operations, $O$,

\begin{quote}

$O\equiv(\cup_{i\in P}$ the operations submitted by
$i)\bigcup(\cup_{x\in V}(w,\epsilon,x,\bot))$

\end{quote}

where $\epsilon$ is a special symbol not used to denote any process,
and $\bot$ is a special value that cannot be written by any process.
The operation $(w,\epsilon,x,\bot)$ is called the \emph{initial write}
of $x$.

\end{definition}

\begin{definition}

\emph{Local order} for process $i$, $<_{iLocal}$,

\begin{quote}

\begin{tabbing}

$<_{iLocal}\equiv$\=$($a total order on $(*,i,*,*))\bigcup$\\
\>$(\forall_{x\in V,o_i\in(*,i,*,*)}\
(w,\epsilon,x,\bot)<_{iLocal}o_i)$

\end{tabbing}

\end{quote}

\end{definition}

\begin{definition}

\emph{Process order}, $<_{PO}$,

\begin{quote}

$<_{PO}\equiv\cup_{i\in P}<_{iLocal}$

\end{quote}

\end{definition}

\begin{definition}

\emph{Writes-to order}, $\mapsto$,

\begin{quote}

$\forall_{(r,i,x,v)\in O}\exists$ unique $(w,j,x,v)\in O$ such that
$(w,j,x,v)\mapsto (r,i,x,v)$

\end{quote}

\end{definition}

These definitions say that the set $O$ includes the operations
submitted by all processes plus an initial write for each variable.
Operations by a single process are totally ordered and are ordered
after all initial writes by local order.  Process order is the union
of all local orders.  Without loss of generality, assume that every
variable has an initial write, and writes are uniquely valued.  As a
consequence of this, for every read there exists exactly one write
that writes-to that read.  Writes-to order is redundant with the
values returned by read operations.  Knowing either one determines the
other, but both are defined for convenience.

\begin{figure}[tp]

\begin{center}

\begin{tabular}{l|l}

$P=\{p_1,p_2\}$                 & $P=\{p_1\}$                     \\
$V=\{x,y\}$                     & $V=\{x\}$                       \\
$O=\{(w,\epsilon,x,\bot),(w,\epsilon,y,\bot)$
                   & $O=\{(w,\epsilon,x,\bot),(w,p_1,x,1),$       \\
\hspace*{1cm}$(w,p_1,x,1),(r,p_1,y,2),$
                   & \hspace*{1cm}$(w,p_1,x,2),(r,p_1,x,1)\}$     \\
\hspace*{1cm}$(r,p_2,x,1),(w,p_2,y,2)\}$                          \\
$(w,\epsilon,x,\bot)<_{PO}(w,p_1,x,1)<_{PO}(r,p_1,y,2)$
                   & $(w,\epsilon,x,\bot)<_{PO}(w,p_1,x,1)<_{PO}$ \\
$(w,\epsilon,y,\bot)<_{PO}(w,p_1,x,1)<_{PO}(r,p_1,y,2)$
                   & \hspace*{1cm}$(w,p_1,x,2)<_{PO}(r,p_1,x,1)$  \\
$(w,\epsilon,x,\bot)<_{PO}(r,p_2,x,1)<_{PO}(w,p_2,y,2)$           \\
$(w,\epsilon,y,\bot)<_{PO}(r,p_2,x,1)<_{PO}(w,p_2,y,2)$           \\
$(w,p_1,x,1)\mapsto(r,p_2,x,1)$ & $(w,p_1,x,1)\mapsto(r,p_1,x,1)$ \\
$(w,p_2,y,2)\mapsto(r,p_1,y,2)$ &                                 \\
                                                                  \\
\multicolumn{1}{c}{(a)}         & \multicolumn{1}{c}{(b)}

\end{tabular}

\end{center}

\caption{Two Executions}

\label{executions}

\end{figure}

An execution defines the operations that were submitted to a memory
system and specifies the externally visible behavior of the memory
system by the writes-to relation.  Now we need to relate the behavior
of the memory system to correctness with respect to a consistency
model.  Consider Figure~\ref{executions}.  Execution (a) corresponds
to a sequentially consistent execution.  From the set of operations,
$O$, and the process order we see that $p_1$ wrote $x$ and then read
$y$, and $p_2$ read $x$ and then wrote $y$.  From the writes-to order
we see that $p_2$ read $p_1$'s write, and $p_1$ read $p_2$'s write.
This corresponds to a sequential order of:

\begin{quote}

$(w,\epsilon,x,\bot)<(w,\epsilon,y,\bot)<(w,p_1,x,1)<(r,p_2,x,1)<(w,p_2,y,2)<(r,p_1,y,2)$

\end{quote}

where $<$ denotes an unnamed total order.  Execution (b), however, is
a little disconcerting.  There is one process.  $p_1$ wrote $1$ to
$x$, then wrote $2$ to $x$, and then read $x$.  Unfortunately, the
read returned the value $1$ from the first write, and not $2$ from the
second.  When we try to create a total order we run into a
contradiction.  If the order is:

\begin{quote}

$(w,\epsilon,x,\bot)<(w,p_1,x,1)<(w,p_1,x,2)<(r,p_1,x,1)$

\end{quote}

then the read does not read from the most recent write, but if the
order is:

\begin{quote}

$(w,\epsilon,x,\bot)<(w,p_1,x,1)<(r,p_1,x,1)<(w,p_1,x,2)$

\end{quote}

then this violates process order.  The important thing to note is that
this does qualify as an execution.  Imagine a computer with out of
order instruction dispatching.  If this dispatching mechanism were
buggy it might accidentally switch the order of a read and write to
the same variable.  Execution (b) exactly models this sort of
phenomenon.  However, it is not likely that this execution will be
deemed correct by any consistency model.  The problems we just saw
with creating a total order also give us a hint about how to define a
consistency model in terms of allowable executions.

\begin{definition}

A \emph{view} is a total order on a set of operations representing one
process' view of the sequence of events within the memory system.

\end{definition}

\begin{definition}

A view is \emph{serial} iff every read returns the value from the most
recent (defined by the order of the view) write to the same variable.

\end{definition}

\begin{definition}

A view is said to \emph{respect} a relation if every edge in the
relation appears in the view.

\end{definition}

\begin{definition}

A relation, $<$, can be \emph{restricted} to a subset of operations,
denoted $<|subset$, which results in a relation containing the set of
edges that are both in $<$ and between two operations in $subset$.

\end{definition}

The notation, SerialView$(<|subset)$, denotes a serial view over the
operations in $subset$ respecting the relation $<|subset$.  Usually,
$subset$ will be defined in terms of operation patterns, or if
$subset$ is the entire set $O$ the shorthand SerialView$(<)$ will be
used.

\end{document}